\documentclass[fleqn,usenatbib]{mnras}

\usepackage{newtxtext,newtxmath}
\usepackage[T1]{fontenc}
\usepackage{multirow}
\DeclareRobustCommand{\VAN}[3]{#2}
\let\VANthebibliography\thebibliography
\def\thebibliography{\DeclareRobustCommand{\VAN}[3]{##3}\VANthebibliography}

\usepackage{graphicx,lipsum,pdflscape}

\usepackage{multirow}
\usepackage{amsmath,siunitx}
\usepackage{subcaption}
\usepackage{hyperref}
\usepackage{amsfonts}

\title[UV analysis of the SMC-Shell region]{UVIT Study of the MAgellanic Clouds (U-SMAC) I. Recent star formation history and kinematics of the Shell region in the North-Eastern Small Magellanic Cloud}

\author[Hota et al.]{
Sipra Hota,$^{1,2}$\thanks{E-mail: sipra.hota@iiap.res.in}
Annapurni Subramaniam,$^{1}$
S. R. Dhanush$^{1,2}$, Maria-Rosa L. Cioni$^{3}$ and Smitha Subramanian$^{1}$
\\
$^{1}$Indian Institute of Astrophysics, $2^{nd}$ Block, Koramangala, Bangalore-560034, India\\
$^{2}$Pondicherry University, R.V.Nagar, Kalapet, Puducherry-605014, India\\
$^{3}$Leibniz-Institut für Astrophysik Potsdam (AIP), An der Sternwarte 16,
D-14482 Potsdam, Germany
}

\date{Accepted XXX. Received YYY; in original form ZZZ}

\pubyear{2024}

\begin{document}

\label{firstpage}
\pagerange{\pageref{firstpage} -- \pageref{lastpage}}
\maketitle

\begin{abstract}
The interactions between the Magellanic Clouds significantly affect the shape and distribution of the young stellar population, particularly in the periphery of the Small Magellanic Cloud (SMC). We present the first far-UV (FUV) map of the north-east SMC-Shell region using the Ultra Violet Imaging Telescope (UVIT) onboard AstroSat. The detected FUV stars are combined with Gaia Early Data Release 3 data to create a FUV--optical catalog of $\sim$ 14,400 stars. FUV--optical colour-magnitude diagrams are used along with isochrones to estimate the stellar ages. The detected stars are formed in multiple episodes. We identified two episodes of star formation ($\sim$ 60 and $\sim$ 260 Myr ago) where the episode at $\sim$ 260 Myr is linked to the recent interaction with the Large Magellanic Cloud (LMC) and the episode at $\sim$ 60 Myr is linked to the pericentric passage of the SMC around our Galaxy. The median proper motion (PM) and velocity dispersion are found to be similar to the SMC main body, indicating that this region has not experienced significant tidal effects. The FUV stellar surface density and the dispersion in PM suggest an extent of the inner SMC in the north-east direction to be around 2.2 deg. We detect arm-like and arc-like structures in the FUV stellar density map, and their kinematics appear to be similar to the SMC main body. These extended outer features are the spatial stellar overdensities formed over multiple episodes of star formation, but without apparent kinematic distinction.
\end{abstract}



\section{Introduction}
The Magellanic System is one of the nearest interacting systems of the Milky Way (MW) Galaxy. This interacting system has four components: the Large Magellanic Cloud (LMC), the Small Magellanic Cloud (SMC), the connecting bridge between the LMC and the SMC, known as the Magellanic Bridge (MB), and the Magellanic Stream (MS). These two galaxies, the LMC and the SMC, together known as the Magellanic Clouds (MCs), are located in close proximity to the MW at a distance of $\sim{50}$ kpc \citep{2014AJ....147..122D} and $\sim{60}$ kpc \citep{2015AJ....149..179D}, respectively. The MS contains only gas and follows the orbits of the MCs \citep[][]{2008ApJ...679..432N,2016ARA&A..54..363D,2020Natur.585..203L}, while the MB comprises stars and gas \citep[][]{Hindman1963ALR,1985Natur.318..160I}. These are signatures of tidal interactions of the MCs with each other and/or with the MW.

Until approximately two decades ago, it was accepted that the MCs orbit our Galaxy and that episodic star formation was attributed to the perigalactic passages and/or the corresponding tidal effects of MW-LMC-SMC interactions \citep[][]{1995ApJ...439..652L,2004AJ....127.1531H}. Subsequently, observational studies using the Hubble Space Telescope (HST) and dynamical simulation \citep{2006ApJ...652.1213K,besla2007magellanic,2023arXiv230615719C} found that the MCs are on their first passage around the MW. Recently, \citet{2023arXiv230604837V} proposed a scenario where the LMC is on its second passage.

The SMC is an irregular gas-rich dwarf galaxy with ongoing star formation. Since its dynamical mass \citep[$\sim 10^9 M_\odot$;][]{van_der_Marel_2008} is about 10 and 100 times less than the LMC and the MW, respectively, its shape and the distribution of the young population are dictated by dynamical interactions \citep[][]{2020MNRAS.498.1034M,10.1093/mnras/staa3857}, especially in the less gravitationally bound outer regions. Tracing the recent star formation history in the outskirts can unlock crucial information about the nature of the interactions, whereas kinematics can guide us to trace the impact of these interactions. 

The SMC-Shell region, located in the outskirts of the north-east (NE) SMC, shows signatures of recent star formation and is part of the outer arm B \citep[][]{1972VA.....14..163D}. \citet[][]{1978A&A....68..193B}, using photographic plates in the B and V bands, found three young ($\sim$ 60 Myr old) star clusters, suggesting that this arm may be formed due to a recent burst of star formation. \citet[][]{1987A&A...182L...8A} also identified the SMC-Shell region associated with the faint outer arm in the NE part of the SMC, using a plate-scanning mechanism on the photographic plates. Using the Magellanic Cloud Photometric Survey (MCPS) data, \citet[][]{2000ApJ...534L..53Z} found that the SMC-Shell region is dominated by upper main-sequence stars aged about 200 Myr. \citet[][]{2019A&A...631A..98M} combined the deep optical data of the Survey of the MAgellanic Stellar History \citep[SMASH;][]{2017AJ....154..199N} and the proper motion (PM) data from the Gaia Data Release 2 \citep[DR2;][]{2018A&A...616A...1G} to probe the SMC-Shell region. Using clusters and classical Cepheids, they estimated that this region has stars with age $\le$ 150 Myr. They concluded that the origin of the SMC-Shell region is most likely due to the hydrodynamic interaction of the SMC with the LMC or the MW but discarded the possibility of a tidal origin. \citet[][]{2022MNRAS.509.3462P} studied 20 star clusters of the SMC-Shell region using the SMASH data and found that the region is extended more along the line of sight ($\sim$ 13.0 kpc) than formerly studied.
  
  As massive young stars are very hot and are better detected and characterised in the ultra-violet (UV) part of the spectrum, a UV-optical study can lead to a better estimation of the recent star formation history and hence, the interaction. In this work, we have used  UV images from the Ultra Violet Imaging Telescope \citep[UVIT;][]{2017JApA...38...28T}. In addition, we have also used the optical and PM data from Gaia Early Data Release 3 \citep[EDR3;][]{2021A&A...649A...1G},  which are the same with respect to the data in Gaia Data Release 3 \citep[DR3;][]{2022arXiv220800211G,2023A&A...674A...1G}. Though the Galaxy Evolution Explorer \citep[GALEX;][]{2005ApJ...619L...1M} has a partial coverage of the SMC in the UV, the stellar population is not quite resolved in these images. As the spatial resolution of the UVIT is three times better than the GALEX, the UVIT images are better suited to study the SMC young stellar population. We have used a UV--optical stellar map of the SMC-Shell region to decipher its recent star formation history, kinematics and probe the details of two structures found in this region.

 In Section~\ref{section: data}, we describe the observations and data used in this analysis. Section~\ref{section:cmd} and Section~\ref{section:density} describe the age estimation and spatial distribution of the FUV stellar population, respectively. Section~\ref{section:diffpop} presents the kinematics of the young population. The spatial distribution of age and median PM are given in Section~\ref{section:age}, whereas Section~\ref{section:mor} deals with the properties of two morphological structures found within the SMC-Shell region. In Section~\ref{section:diss}, we discuss our results, and we summarise and conclude our work in Section~\ref {section:con}.

\section{DATA sets (UVIT \MakeLowercase{and} G\MakeLowercase{aia})}
\label{section: data}
\begin{figure*}
	\includegraphics[width=\textwidth]{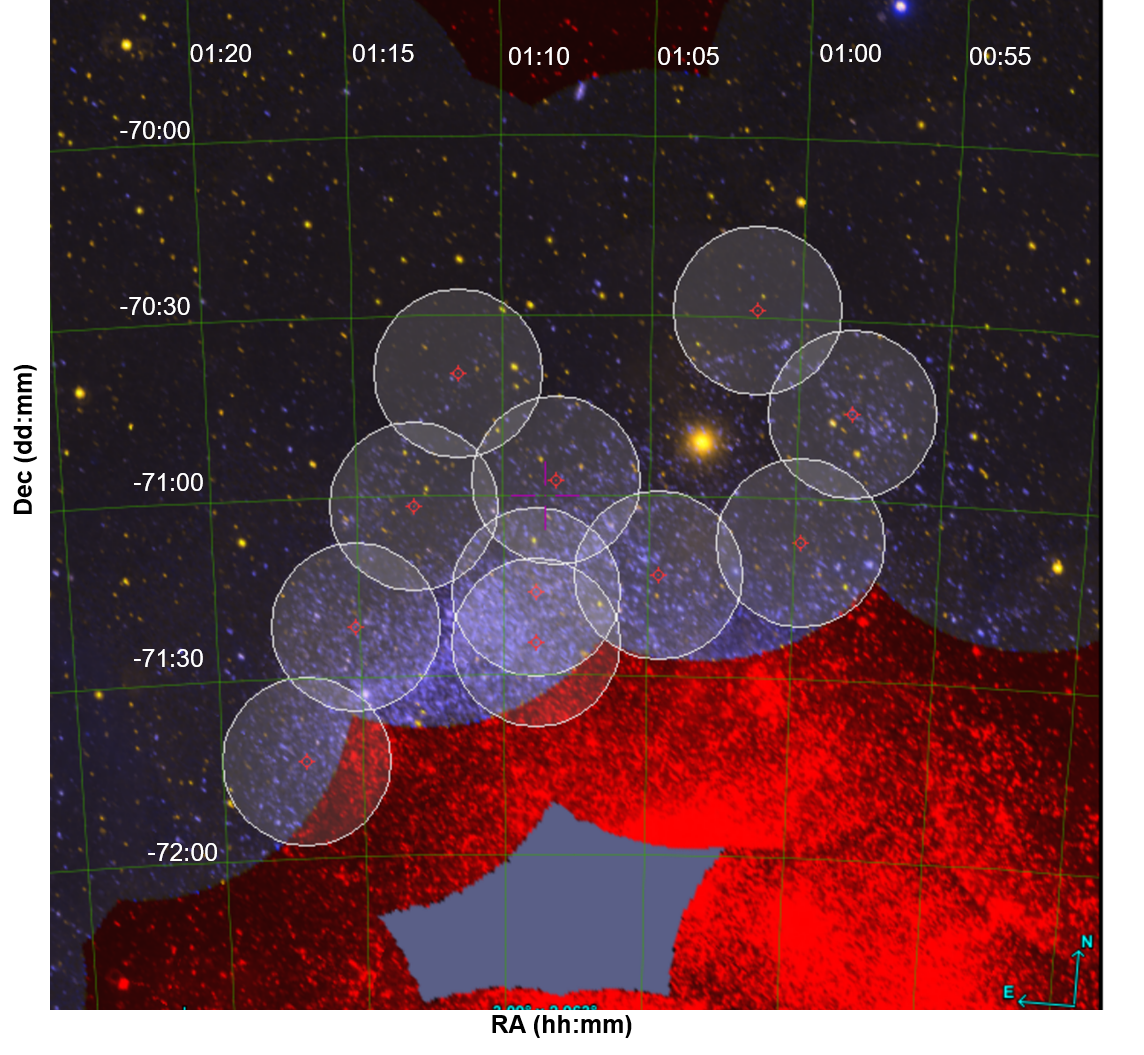}
    \caption{The SMC-Shell region: The GALEX GR6/7 data \citep{2017ApJS..230...24B} are shown as the background, and the fields of the SMC-Shell region, observed using the UVIT/AstroSat, are represented by white circles. This background GALEX image is a 2-color composite image where blue and red correspond to GALEX FUV and GALEX NUV, respectively, highlighting the overlap between the two. The region around the Galactic globular cluster NGC 362 (RA$=01^{\mathrm{h}}$:$0.3^{\mathrm{m}}$, Dec$=-70^{\circ}$:$50^{\prime}$) is excluded from the observing plan.}

    \label{fig:Galex_bg_UVIT_obs_fields}
\end{figure*}

The SMC-Shell region was imaged by the UVIT onboard the AstroSat, the first Indian Space Observatory, as shown in Fig.~\ref{fig:Galex_bg_UVIT_obs_fields}.
 
The UVIT is a twin telescope with a spatial resolution of \ang{;;1.4} and a field of view of 0.5 deg. One telescope is for the far-UV (FUV: 130 -- 180 nm), and the second one is for the near-UV (NUV: 200 -- 300 nm) and the visible (VIS: 350  --  550 nm) observations. The FUV and NUV channels are used to obtain the photometric data, and the VIS channel is used for drift correction due to the spacecraft \citep[][]{2017JApA...38...28T}. In this work, we have obtained FUV images in the F148W (1481 $\pm$ 250 \AA) filter for eleven fields of the SMC-Shell region. Observation details are given in Table~\ref{tab:data_table}. More information about the UVIT calibration and instrumentation can be found in \citet[][]{2012SPIE.8443E..1NK,subramaniam2016orbit,2017AJ....154..128T}. 
\begin{table}
	\centering
	\caption{Observational details of the eleven fields of the SMC-Shell region, observed by using the UVIT/AstroSat. RA and Dec are the center coordinates of the observed fields.}
	\label{tab:data_table}
	\begin{tabular}{lccr} 
	\hline
     Field  & t$_{\mathrm{exps}}$ (sec) & RA (deg) & Dec (deg)\\
    \hline
    Field 1& 2406.4 & 14.55 & $-70.76$ \\
    Field 2& 2405.7 & 14.94 & $-71.12$\\
    Field 3& 2373.2 &	15.37 & $-70.48$\\
    Field 4& 2371.3 &	16.16 &	$-71.22$\\
    Field 5& 2366.2 & 17.05 & $-70.96$\\
    Field 6& 2208.1 &	17.87 &	$-70.66$\\
    Field 7& 1533.6 &	18.26 &	$-71.03$\\
    Field 8& 2373.7 &18.79& $-71.36$\\
    Field 9&2375.1 &19.25& $-71.73$\\
    Field 10& 1903.6 &17.22& $-71.27$\\
    Field 11&1902.5 &17.22& $-71.41$\\
    \hline	
	\end{tabular}
\end{table}

\begin{figure}
  \includegraphics[width=\columnwidth]{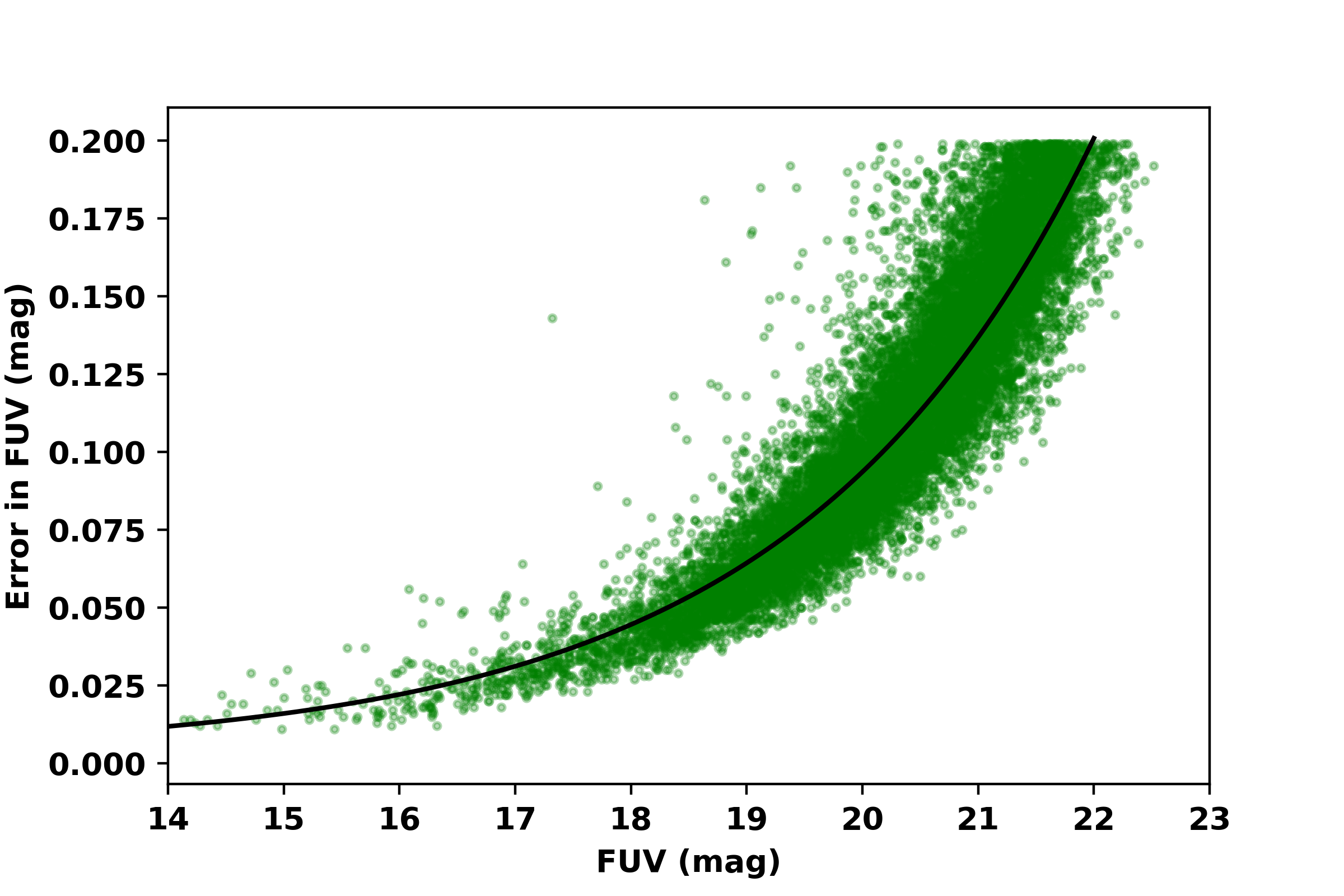}
    \caption{The PSF-fit error as a function of the FUV magnitude for the observed sources of the SMC-Shell region. The line plot is the median error of the F148W filter.}
    \label{fig:Mag_Mage}
\end{figure}

The science-ready images were made from the level 1 data using the CCDLAB pipeline \citep[][]{2017PASP..129k5002P,2021JApA...42...30P}, including astrometry. The photometry of the FUV images was done by performing the point-spread function (PSF) photometry using the DAOPHOT package of IRAF \citep[Image Reduction And Analysis Facility:][]{1987PASP...99..191S}. To obtain the FUV magnitudes of detected stars, we carried out the following steps. First, we located stars in the science-ready images using the DAOFIND task in the IRAF and then performed aperture photometry using the PHOT task. Further, we selected isolated and bright stars in the images to construct the model PSF using the PSTSELECT task. The average PSF of all the stars in the eleven images varied from $\ang{;;1.2}$ to $\ang{;;1.6}$. We fitted the model PSF to all the detected stars in the image using the ALLSTAR task to obtain the PSF-fitted magnitudes. The PSF magnitudes were converted to aperture photometry scale using the PSF correction. Further, we considered aperture correction, using the curve of growth analysis by choosing isolated bright stars in the field, followed by saturation correction. The details of saturation correction are described in \citet{2017AJ....154..128T}. The F148W magnitudes were converted into the AB magnitude system using the zero-point (ZP) given in \citet{2020AJ....159..158T}. More information about the PSF photometry on UVIT observed fields can be found in \citet[][]{2022MNRAS.514.1122S}. The PSF-fit error as a function of the estimated magnitudes in the F148W filter for all the FUV detected sources is shown in Fig.~\ref{fig:Mag_Mage}. For a maximum error of 0.2 mag, we detect sources up to 22 mag and these are considered for further analysis. We combined all the observed fields to create a catalog of $\sim$ 16000 FUV sources in the SMC-Shell region after taking care of data duplication of the overlapping fields based on signal-to-noise criteria. We note that stars brighter than 15 mag in FUV ($\sim$ 20 stars) might be affected by uncorrected saturation and therefore might be slightly brighter than the estimated magnitude.

The detected FUV sources were cross-matched with the Gaia EDR3 data \citep[][]{2021A&A...649A...1G} within a separation of $\ang{;;1}$, and we considered only stars with RUWE ( Renormalised Unit Weight Error) $< 1.4$ \citep[][]{LL:LL-124}. A {$3\sigma$} cut-off is applied to the mean value of the parallax ( parallax = 0.011 mas) and the PM in the Right Ascension (RA) and the Declination (Dec) ($\mu_\alpha cos\delta$ = 0.90 mas yr$^{-1}$, $\mu_\delta = -1.21$ mas yr$^{-1}$) of the detected FUV stars to obtain the most probable members in the SMC-Shell region. The final FUV catalog of the SMC-Shell region has $\sim$ 14400 stars, along with the optical photometry and the PM values from Gaia EDR3 (Table \ref{tab:catalog}).
\begin{table*}
	\centering
	\caption{UV catalog of the SMC-Shell region. The Gaia EDR3 source\_id, spatial coordinates, and magnitude in the F148W filter along with the fit error in magnitude are listed in columns 1 to 5 respectively. The full table is available as online material.}
	\label{tab:catalog}
	\begin{tabular}{lcccr} 
	\hline
    Gaia EDR3 Source\_id & RA (deg) & Dec (deg) & FUV (mag) & error (mag)\\
    \hline
    4690731730417825920 & 18.12019792
& $-70.85794556$ & 19.581 & 0.081\\
    4690734754072739456 & 17.84952625 & $-70.83735222$ & 16.880 & 0.018 \\
    4690734891513810432 & 17.97096000 & $-70.83903333$ & 17.787  & 0.027\\
    4690795742608440192 & 15.29432708 & $-71.23031833$ & 21.119 & 0.106\\
    4690794716120011904 & 15.36130083 & $-71.25074083$ & 21.446 & 0.117\\

    \hline	
	\end{tabular}
\end{table*}

\section{Colour Magnitude Diagram}
\label{section:cmd}
\begin{figure}  
          \centering
          \includegraphics[width=\columnwidth,trim={0 0 0 0},clip]{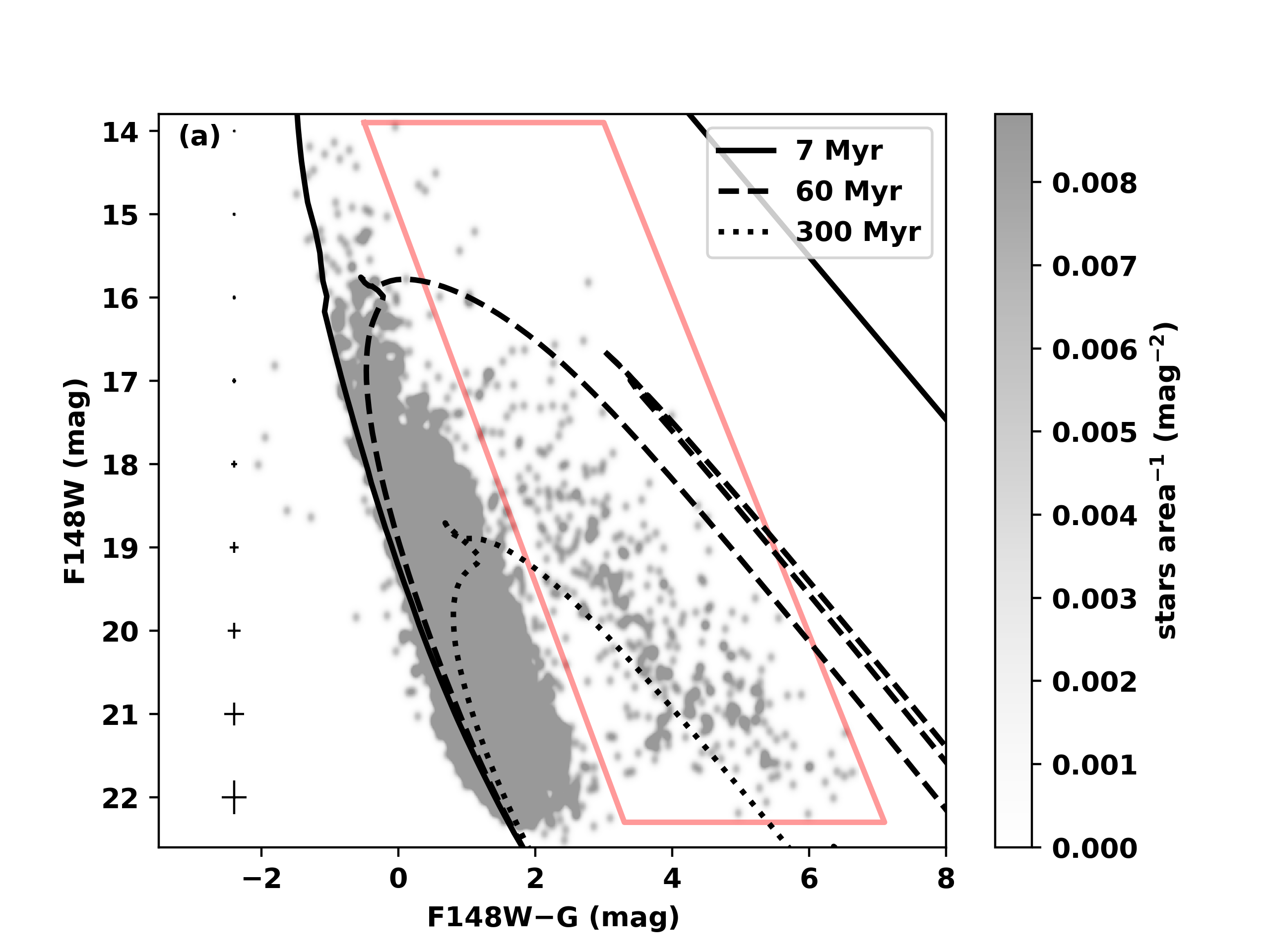}
          \\
           \includegraphics[width=\columnwidth,trim={0 0 0 1.0cm},clip]{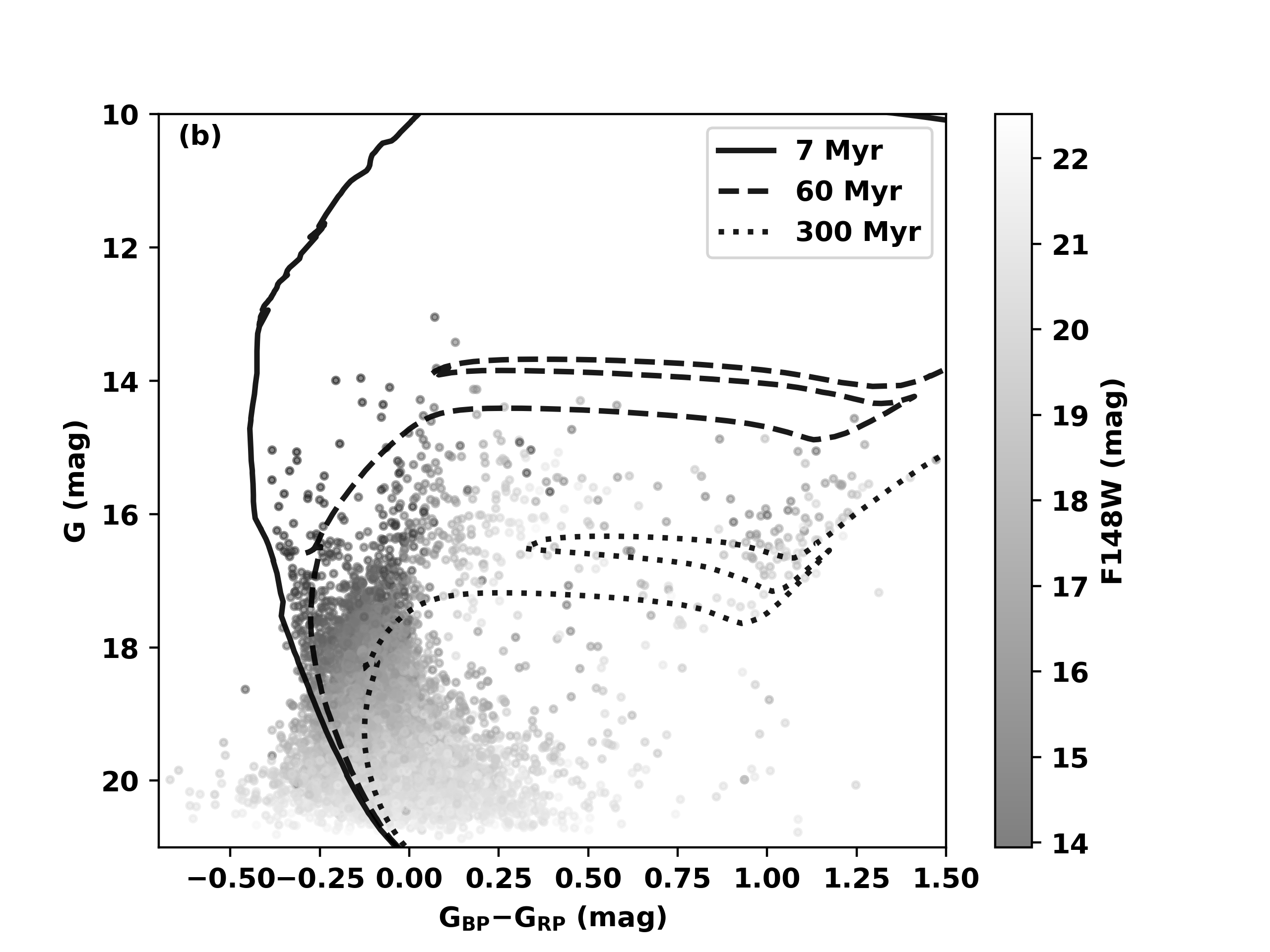}
        
          \caption{(a) FUV--optical CMD (KDE with a bandwidth of 0.05 mag) with the error bars (median value) shown in black colour on the left side and a red polygon denoting the region of the sub-giant branch. (b) Optical CMD of the SMC-Shell region. The colour bars of (a) and (b) represent the number of stars area$^{-1}$ (area = 10$^{-4}$ mag$^2$) and FUV (mag) respectively.}\label{fig:cmd}
 \end{figure}
We used all the stars in the catalog of the SMC-Shell region (Table \ref{tab:catalog}) to construct the colour-magnitude diagram (CMD). We used FUV and optical photometric data to create the FUV--optical (F148W vs (F148W--G)) CMD (Fig.~\ref{fig:cmd}a) and the optical (G vs G$_{\mathrm{BP}}$--G$_{\mathrm{RP}}$) CMD (Fig.~\ref{fig:cmd}b). In Fig.~\ref{fig:cmd}a, we present a kernel density estimation (KDE) applied to the colour-magnitude surface. The bandwidth used for this KDE was set to 0.05 mag. This diagram is aimed to help in identifying the densely populated evolutionary features. We used the Padova-PARSEC isochrones \citep[][]{2012MNRAS.427..127B} with an age step size of 5 Myr to overlay on the CMDs. We adopt the distance, reddening, and metallicity values previously used by \citet[][]{2019A&A...631A..98M}, which are: a distance modulus of 18.96 mag \citep[][]{2015AJ....149..179D}, a metalicity of Z = 0.002 dex \citep[][]{1992ApJ...384..508R,2008A&A...488..731R,2017A&A...608A..85L} and a mean colour excess of $E(B-V) = 0.05$ mag \citep[][]{2011AJ....141..158H}. When we overlaid the isochrones in the FUV--optical CMD, we had to reduce the colour excess value to 0.03 mag, which is within the error margin of the literature reddening value, to best visually overlay the isochrones onto the data. On the other hand, we adopted the same extinction value as in the literature for the optical CMD. The difference in reddening (0.02 mag) between those adopted in the FUV--optical and the optical CMDs is very small. In the optical CMD, the blue loop stars and the sub-giant stars overlap, resulting in an ambiguous age estimation from the main-sequence turnoff. But in the FUV--optical CMD, the large colour-magnitude range introduces a sufficient separation between the blue loop and the sub-giant stars. Therefore, the FUV--optical CMD is better suited for estimating the age range of young stars. From Fig.~\ref{fig:cmd}a and Fig.~\ref{fig:cmd}b, the sub-giant branch is clearly seen in the FUV--optical CMD (within a polygon with coordinates at (F148W$-$G, F148W) $=$ [($-$0.5, 13.9), (3, 13.9), (7.1, 22.3), (3.3, 22.3), ($-$0.5, 13.9)] mag when compared to the optical CMD (at G $\approx$ [14, 17] mag and G$_{\mathrm{BP}}$--G$_{\mathrm{RP}}\approx$ [0, 1.5] mag). To guide the reader with the stellar evolutionary phases detected in the CMD, we have overlaid a few isochrones to demonstrate the extended turn-off and the sub-giant branch in the FUV magnitude axis. The overlaid isochrones suggest that most of the star formation is likely to have happened between 60 -- 300 Myr ago and a probable recent event at $\sim$ 7 Myr. We note that the brightest stars in the CMD may be brighter due to uncorrected saturation, but the younger age estimations are unlikely to be affected by this.

\section{Density map of the FUV population}
\label{section:density}

\begin{figure}
  \includegraphics[width=\columnwidth]{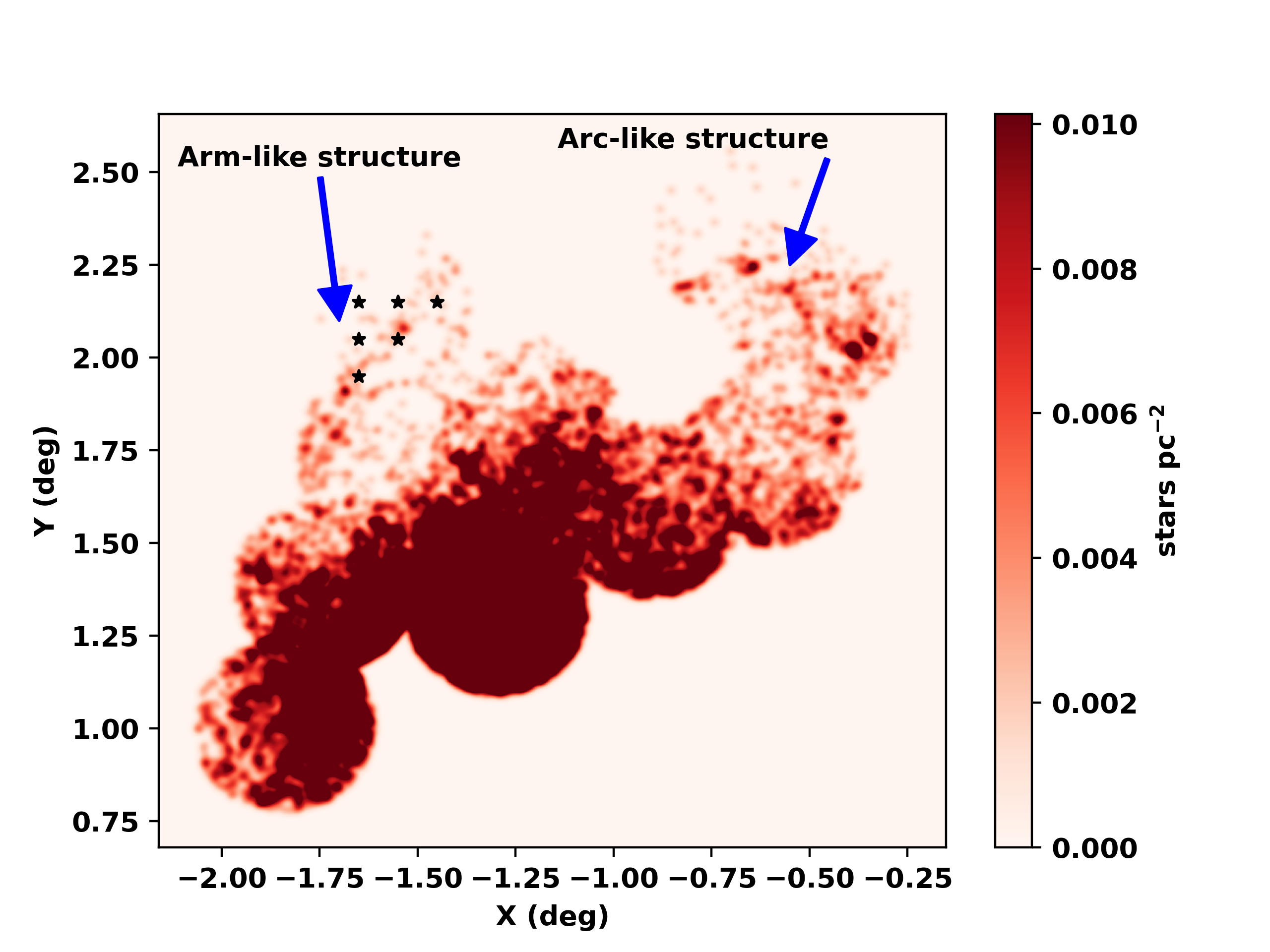}
    \caption{The surface density plot of the SMC-Shell region. The colour bar represents the number of stars pc$^{-2}$. Blue arrows indicate the arm-like and arc-like structures. The six black points are the center of the bins located on the arm-like structure which are shown within the blue polygon in Fig.~\ref{fig:den}b.}
    \label{fig:cartesian map}
\end{figure}

    \begin{figure}  
        \centering
        \includegraphics[width=\columnwidth,trim={0 0 0.7cm 0},clip]{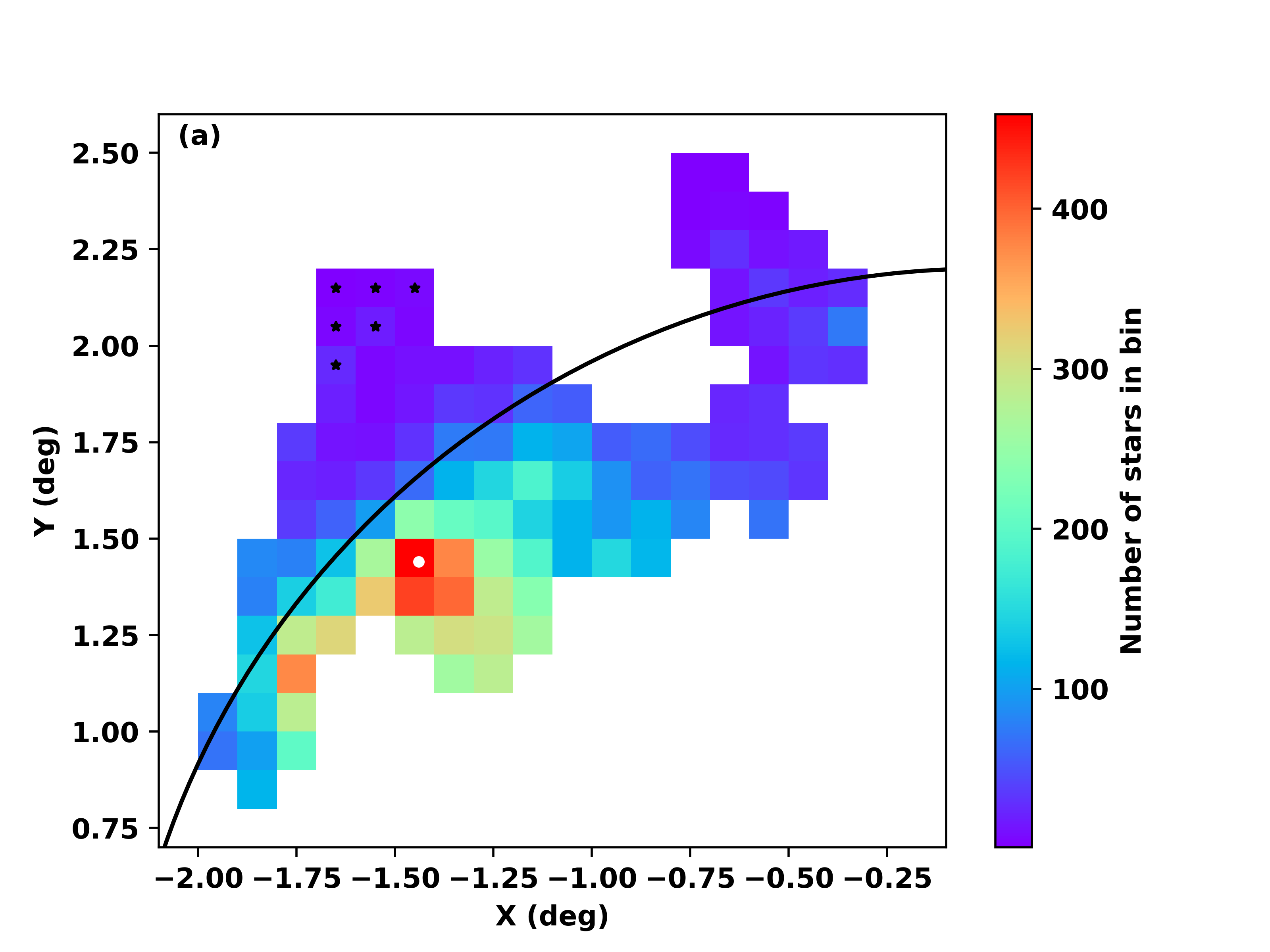}
         \\
        \includegraphics[width=\columnwidth,trim={0 0 0 1.0cm},clip]{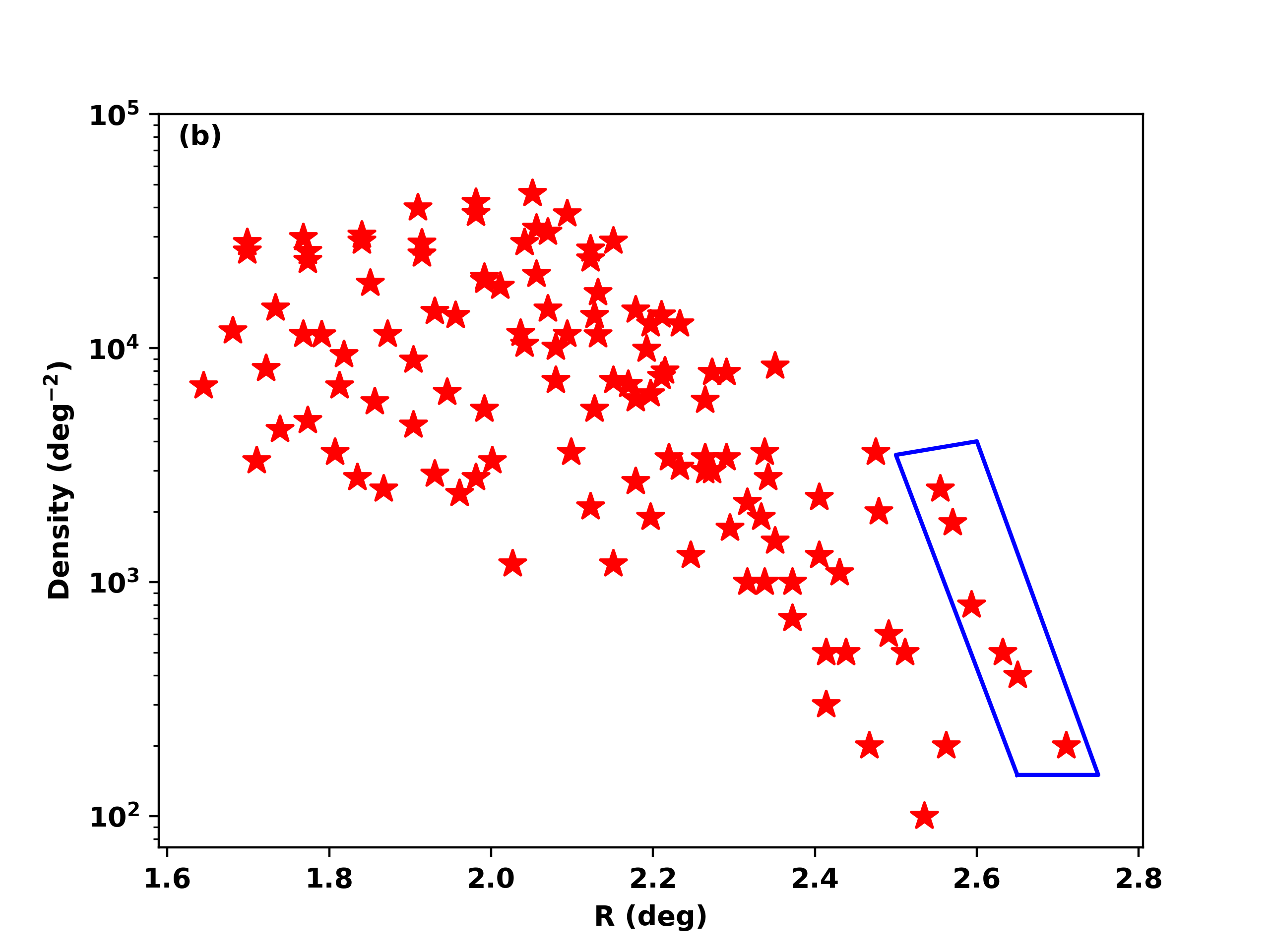} \\
        \caption[]{(a) The spatial distribution of FUV stars of the SMC-Shell region. The colour bar shows the number of stars in each bin (area of bin = 0.01 deg$^2$). The black arc represents a segment of a circle with a radius of 2.2 deg centered at the SMC optical center. The white dot represents the location of star cluster HW64. (b) The radial variation of the number density of FUV stars from the SMC center. The points in the blue polygon in Fig.~\ref{fig:den}b and marked as stars Fig.~\ref{fig:den}a are located on the arm-like structure shown in Fig.~\ref{fig:cartesian map}.}
        \label{fig:den}
    \end{figure}
We have determined zenithal equidistant projection of the SMC-Shell region where we followed the X and Y conversion as defined in \citet[][]{2001AJ....122.1807V} with the SMC optical center at $\alpha_{\mathrm{SMC}} = 00^{\mathrm{h}} 52^{\mathrm{m}} 12^{\mathrm{s}}.5$ and $\delta_{\mathrm{SMC}}$ =$\ang{-72}$ $\ang{;49}$ $\ang{;;43}$\citep[J2000;][]{1972VA.....14..163D}. To understand the spatial distribution of FUV stars in the SMC-Shell region, we used KDE, where convolution of stellar distribution with a Gaussian kernel was performed. Using a kernel size of 10 pc, we could identify morphological structures.  The same kernel size was considered in previous studies on the MCs by \citet{2017ApJ...849..149S,2018ApJ...858...31S,2022MNRAS.512.1196M}. We have visually identified two structures within this region; one is an arm-like structure, and another is an arc-like structure, shown in Fig.~\ref{fig:cartesian map}. These two structures were noticed previously by \citet[][]{2019A&A...631A..98M} using main-sequence stars brighter than 21 mag in the g band (right panel of their Fig. 3 and Fig. 4) obtained from the SMASH data. The details of these two structures are explained in Section~\ref{section:mor}.

We performed a spatial binning with the bin size of 0.1 deg along the X and Y directions and estimated the number of stars per bin. Here, we considered only bins that fully fall within the FUV coverage (118 such bins). The spatial distribution of FUV stars is shown in Fig.~\ref{fig:den}a. In Fig.~\ref{fig:den}a, we identified the densest bin with center at (X, Y) $=$ ($-1.45$, 1.45) deg where a young cluster HW64 of age $\sim$ 30 Myr \citep[][]{2022MNRAS.509.3462P} is located. We identified the positions of clusters within the SMC-Shell region, which had been previously examined by \citet[][]{2022MNRAS.509.3462P}. We observed that these clusters lack a significant population of FUV stars, and the FUV--optical CMDs do not offer substantial insights for further investigation. Consequently, we opted not to explore these detected clusters in our study. We note that the spatial distribution of young stars is not uniform, and there is a gradient that decreases radially outward (Fig.~\ref{fig:den}a).

We calculated the FUV stellar surface density of bins with area of 0.01 deg$^2$ and their radial distance from the SMC center (Fig.~\ref{fig:den}b).
We note a nearly flat FUV stellar surface density between 1.6 to 2.2 deg, where the density range is found to be between a few thousand to a few tens of thousands of stars deg$^{-2}$ and the same does not occur for all the position angles. The density starts to decline, from $\sim$ 2.2 deg and falls below 1000 stars beyond $\sim$ 2.4 deg. The radius corresponding to 2.2 deg is shown in Fig.~\ref{fig:den}a as a black curve. This suggests that the area within this radius covers nearly all the higher density regions within the limitation of spatially incomplete observations in the NE of the SMC (X $>-1$ deg) and hence could be considered as an extent of the inner SMC within the limited observed part of the NE SMC. Regions beyond this radius could be considered as the outer SMC because they have a declining stellar density. The six outermost points showing a higher stellar density more than 100 FUV stars deg$^{-2}$ are identified in the arm-like structure and shown in Fig.~\ref{fig:cartesian map} and Fig.~\ref{fig:den}a. The arc-like structure appears to start within the inner SMC and extends to the outer SMC, whereas the arm-like structure seems entirely co-located in the outer SMC (Fig.~\ref{fig:den}a).

\section{Kinematics of different age groups}
\label{section:diffpop}
    \begin{figure*}  
        \centering
        \includegraphics[width=\columnwidth,trim={0.5cm 0 1.0cm 0},clip]{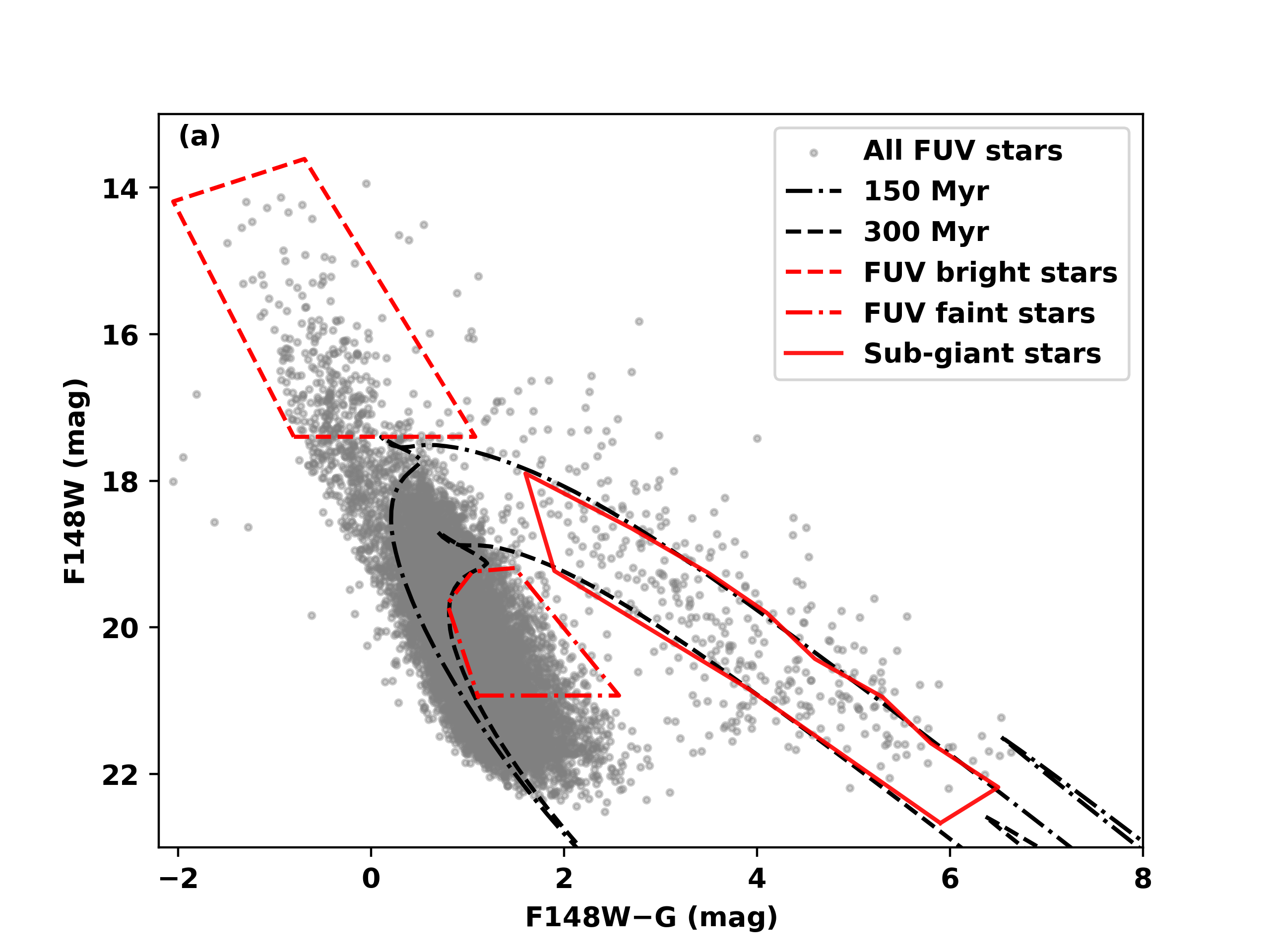}
        \includegraphics[width=\columnwidth,trim={0.35cm 0 0 0},clip]{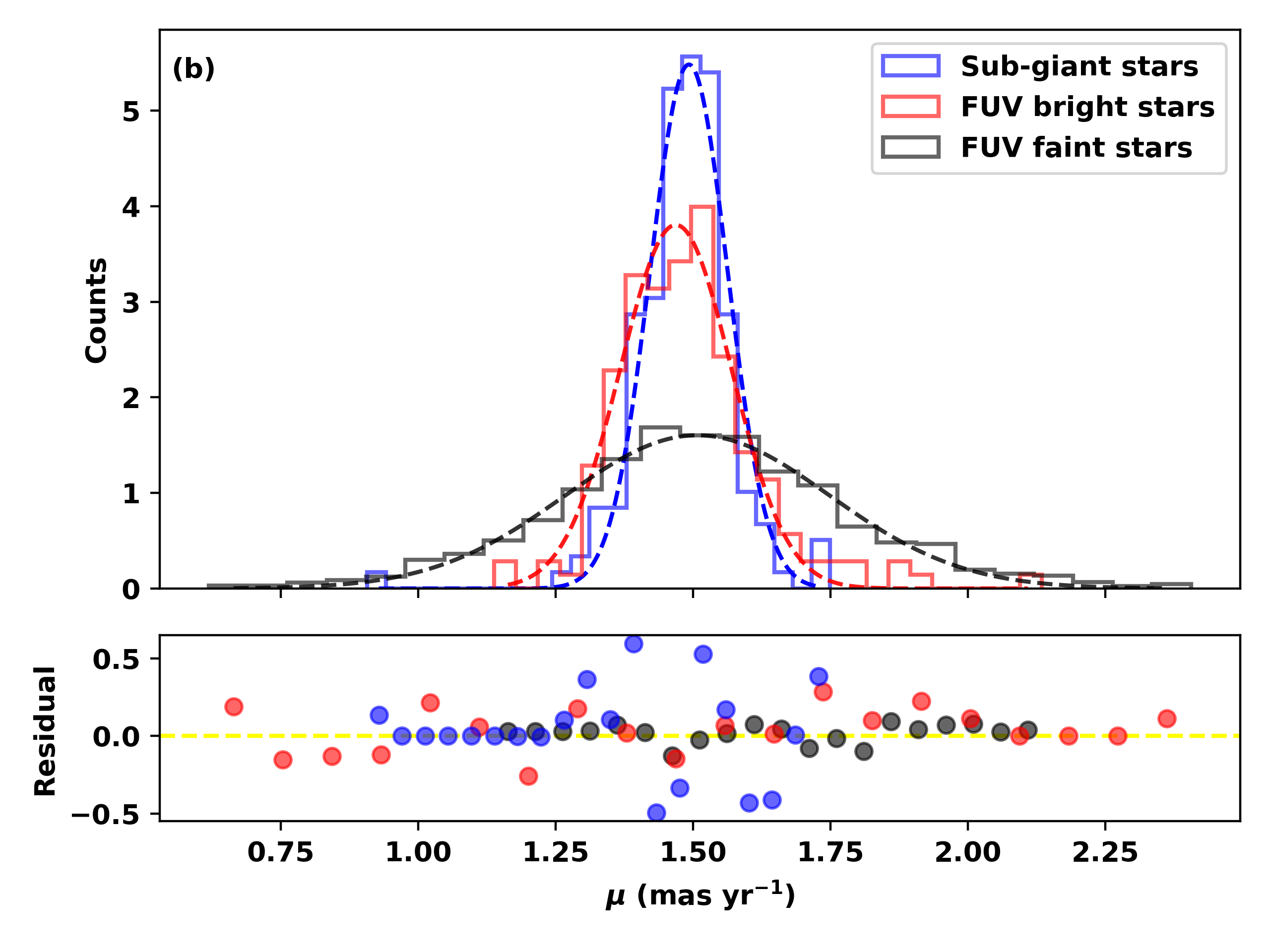} \\
        \includegraphics[width=\columnwidth,trim={0.1cm 0 1.0cm 0.5cm},clip]{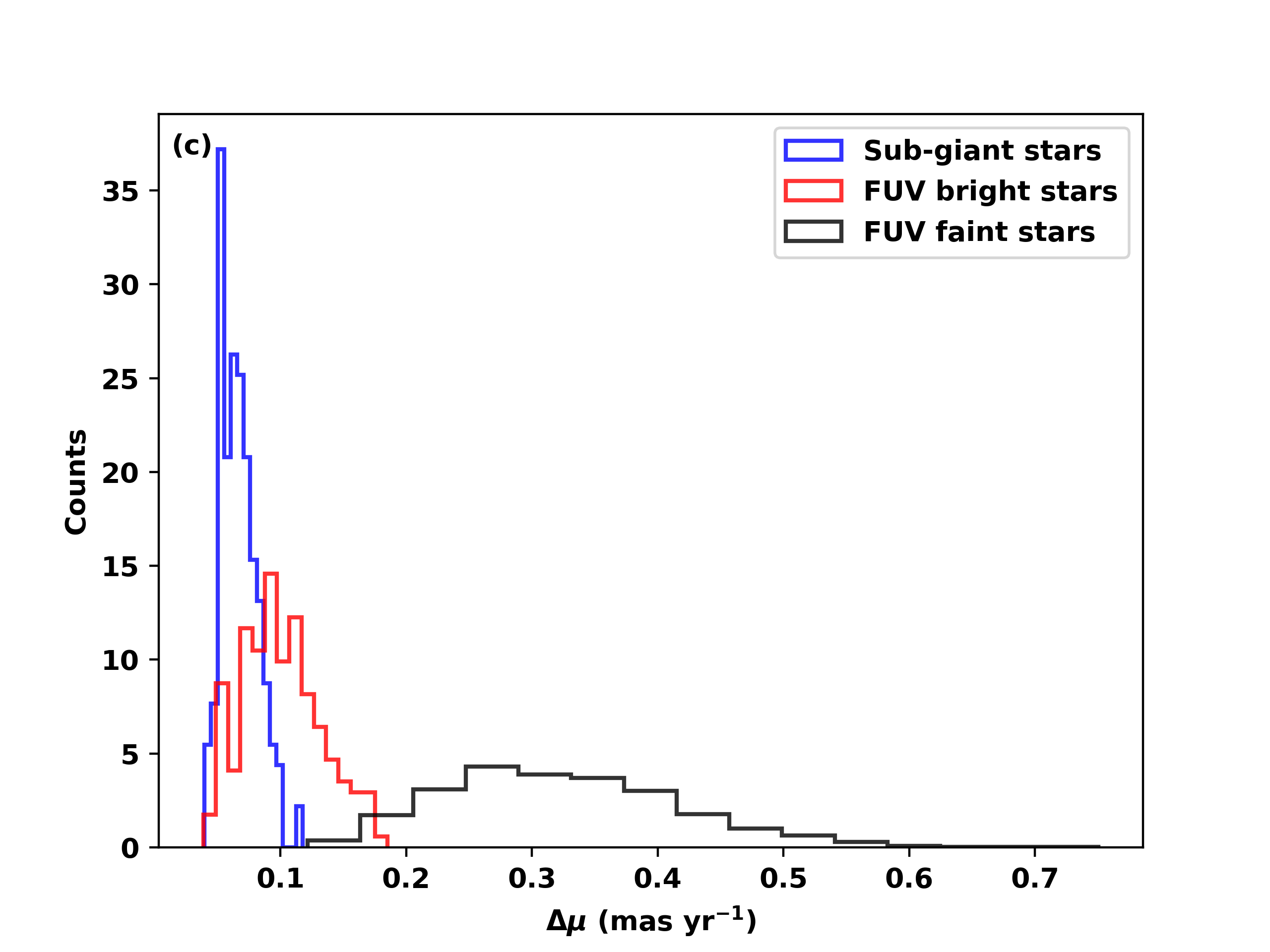}
        \includegraphics[width=\columnwidth,trim={0.1cm 0 1.0cm 0.5cm},clip]{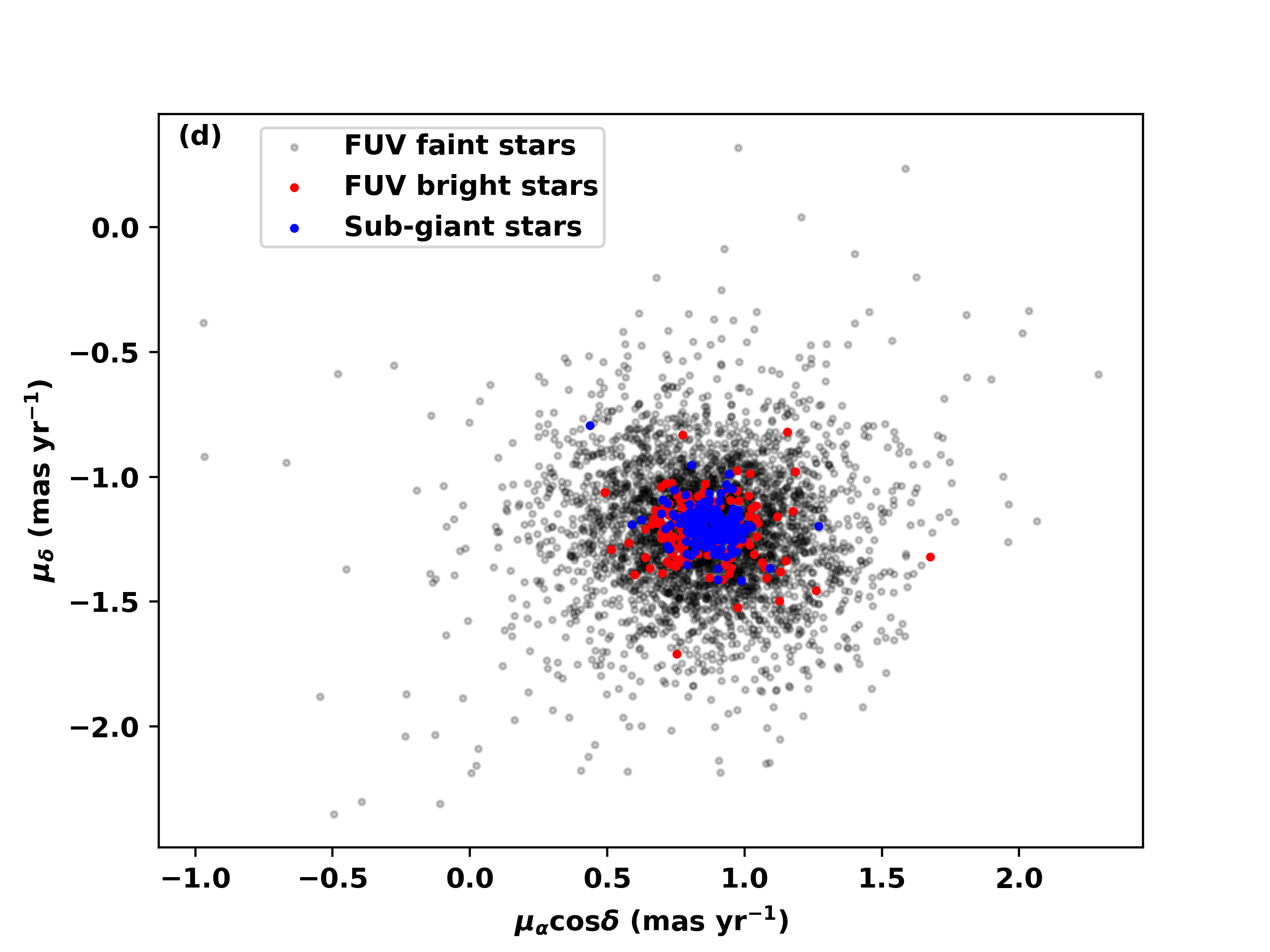}\\
        
        \caption[]{(a) Selection of different populations using the FUV--optical CMD: FUV bright (red dashed polygon), FUV faint (red dash-dotted polygon), and sub-giant (red solid polygon) stars. (b) the Gaussian fit of the histogram of the PM with residuals, (c) the distribution of measurement error in PM, and (d) the VPD of the three different populations.}
    \label{fig:Kinematics of different populations}
    \end{figure*}

As the outer regions are affected more by tidal effects due to the recent interaction between the MCs,  it will be interesting to probe the kinematics of stars born before and after the interactions in this part of the SMC. To differentiate the kinematics of FUV stars formed before and after the recent LMC-SMC interaction, we defined three age groups from the FUV--optical CMD as shown in Fig.~\ref{fig:Kinematics of different populations}a. Stars with FUV magnitudes brighter than 17.4 mag and younger than 150 Myr in the main-sequence of the CMD are defined as the FUV bright population, and stars with the FUV magnitude between 19 to 21 mag in the main-sequence are defined as the FUV faint population, as the majority of this population are stars older than 300 Myr. The third group is the sub-giant stars, aged between 150 to 300 Myr and fainter than 18 mag.

First, we applied a 3$\sigma$ cut-off to the PM of these populations to remove outliers, then we created the PM distributions (Fig.~\ref{fig:Kinematics of different populations}b), which we fitted with Gaussian and estimated the values of mean and standard deviation of the PM distribution, along with median values of PM and median values of error in PM of each group as shown in Table~\ref{tab:data_table_1}. The error in PM of each star of these three populations is estimated by using ${\mathrm{\Delta\mu_{\alpha}cos\delta}}$ and ${\mathrm{\Delta\mu_\delta}}$ from Gaia, and the distributions of these errors are shown in Fig.~\ref{fig:Kinematics of different populations}c.

\begin{table}
	\centering
	\caption{For each of the three FUV populations listed in column 1, mean and standard deviation values of the PM distribution with their fit error (from the covariance matrix) are listed in columns 2 and 3, respectively. The median values of PM and the median value of error in PM are listed in columns 4 and 5, respectively.}
    \label{tab:data_table_1}
    \begin{tabular}{ |p{1.9cm}|p{1.1cm}|p{1.1cm}|p{1.1cm}|p{1.1cm}|  }
	\hline
    Defined population  & $\mu_{\mathrm{mean}}$ \newline(mas yr$^{-1}$) & $\sigma_\mu$ \newline(mas yr$^{-1}$) &$\mu_{\mathrm{median}}$\newline (mas yr$^{-1}$) & $\Delta\mu_{\mathrm{median}}$ \newline(mas yr$^{-1}$)\\
    \hline
    FUV faint stars  & 1.508 $\pm$ 0.006 & 0.241 $\pm$ 0.006 & 1.508 &  0.315  \\
    FUV bright stars  & 1.468 $\pm$ 0.003 & 0.101 $\pm$ 0.003 & 1.471  & 0.097 \\
    Sub-giant stars & 1.492 $\pm$ 0.003& 0.070 $\pm$ 0.003 &  1.490 & 0.065 \\
    \hline	
	\end{tabular}
\end{table}
We converted the standard deviation in PM to velocity dispersion using the following formula.
\begin{equation} \label{eq1}
V  = 4.7 \mu D \\
\end{equation}
where V is the transverse velocity in km s$^{-1}$ and D is the distance in parsec (pc).

We obtained the velocity dispersion of the FUV bright and the sub-giant stars as 29.67 $\pm$ 0.88 km s$^{-1}$ and 20.76 $\pm$ 0.88 km s$^{-1}$, respectively. The velocity dispersion of the FUV faint stars is 70.82 $\pm$ 1.76 km s$^{-1}$, which is very large compared to the velocity dispersion of the other two groups. Fig.~\ref{fig:Kinematics of different populations}d is a Vector Point Diagram (VPD) of the three groups. This plot illustrates that the FUV bright and sub-giant stars appear compact when compared to the FUV faint stars. We note that the median value of error in PM, 0.315 mas yr$^{-1}$ is relatively higher than the standard deviation of 0.241 mas yr$^{-1}$ of PM distribution for this population. Therefore, the relatively large velocity dispersion estimated for the age group $>$ 300 Myr is likely to be due to the large measurement error in PM. However, the FUV bright and sub-giant stars have similar peak and standard deviation values in PM, along with a small difference in velocity dispersion. This leads us to conclude that the stars younger than 150 Myr (FUV bright stars) and stars with age between 150 -- 300 Myr (sub-giants) are kinematically similar. 
As the kinematic dispersion is used as a probe to check for interaction induced by tidal heating of the older stellar population in the outer SMC, a consistent dispersion points to an undetectable tidal heating within the data accuracy.

\section{Spatial Distribution of Age and Proper Motion }
\label{section:age}

In order to assess the age and kinematics of FUV stars comprehensively, we obtained the age range and median PM for each individual spatial bin within the SMC-Shell region. We considered the same binning as mentioned in Section~\ref{section:density} for the age map, but we excluded bins (regions) containing less than 15 FUV stars to get a reliable estimate of the age range. For each of the 126 individual bins, we created FUV--optical CMDs. These CMDs bear the signatures of the episodes of star formation experienced by the region in the form of evolutionary features in the CMD. As these CMDs cannot be considered as originating from a simple star population, we cannot use quantitative isochrone fitting methods to estimate ages. Therefore, we utilize the presence of (1) sub-giants and (2) turn-off features in the CMDs to identify the dominant population, and Padova-PARSEC isochrones are visually overlaid on them to derive their ages.
We did not come across any requirement to alter the assumed values of extinction and distance (Section~\ref{section:cmd}), as the overlaid isochrones matched the stellar patterns in the CMD well.

In most of the regions, we were able to detect up to three episodes of star formation based on the features in the CMD (for example, Fig.~\ref{fig:parm}b and Section~\ref{subsec:arm}). These suggest that in general, the SMC-Shell region experienced star formation in episodes. We combined the identified ages of star formation in the individual regions to understand the overall picture of star forming episodes in the SMC-Shell region. 

As the ages are estimated by a visual overlay of isochrones on the CMD features, we overlaid isochrones with a range in age to identify how much deviation can be detected visually for the age range considered in this study.  We found that it is generally not possible to differentiate isochrones with age within the 10\% range (see, Fig.~\ref{fig:arm_appendix} in Appendix~\ref{append: appendix}). A similar method for estimating the error in age when the fitting was done visually was used by \citet{2014..Piatti..Age}.

\begin{figure*}
    \centering
        \includegraphics[width=0.8\textwidth]{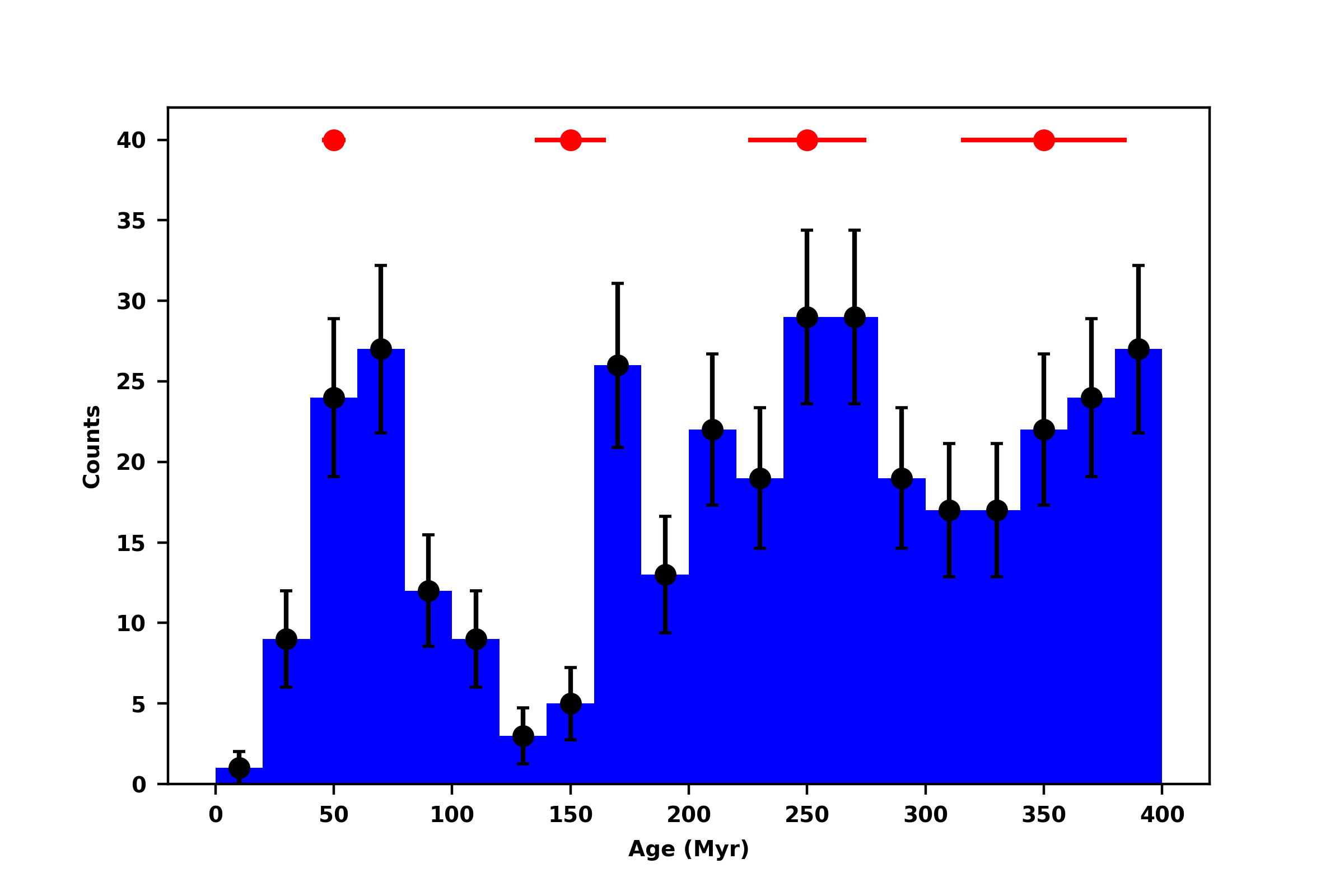}
        \caption{The histogram of the star formation events of the SMC-Shell region. The vertical error bars represent Poissonian errors, while the horizontal error bars at 50 Myr, 150 Myr, 250 Myr, and 350 Myr indicate approximately $\sim 10\%$ errors in the estimated ages.}
    \label{fig:hst_age}
\end{figure*}

The age distribution, as shown in Fig.~\ref{fig:hst_age}, indicates a few peaks in star formation, pointing to the episodic star formation experienced by the region. We used a bin size corresponding to 1-sigma error (20 Myr) at 200 Myr. We note a 
peak between 40 -- 80 Myr and another between 240 -- 280 Myr. The peak noticed at $\sim$ 170 Myr may not be significant and may only suggest a ramping down of star formation between 150 -- 200 Myr. There is a dip between 100 -- 160 Myr, prior to a peak at $\sim$ 60 Myr, indicating a recent burst. The peak at $\sim$ 390 Myr is not regarded as a significant star formation peak as stars with age $\sim$ 390 Myr have a large photometric error. In summary, two noticeable star formation episodes can be traced between 40 -- 300 Myr. The episode of star formation at 240 -- 280 Myr is likely due to the last LMC-SMC interactions \citep[]{1985PASA....6..104M,2012MNRAS.421.2109B,2022ApJ...927..153C}.

\begin{figure*}
 \centering
	\includegraphics[width=0.84\linewidth]{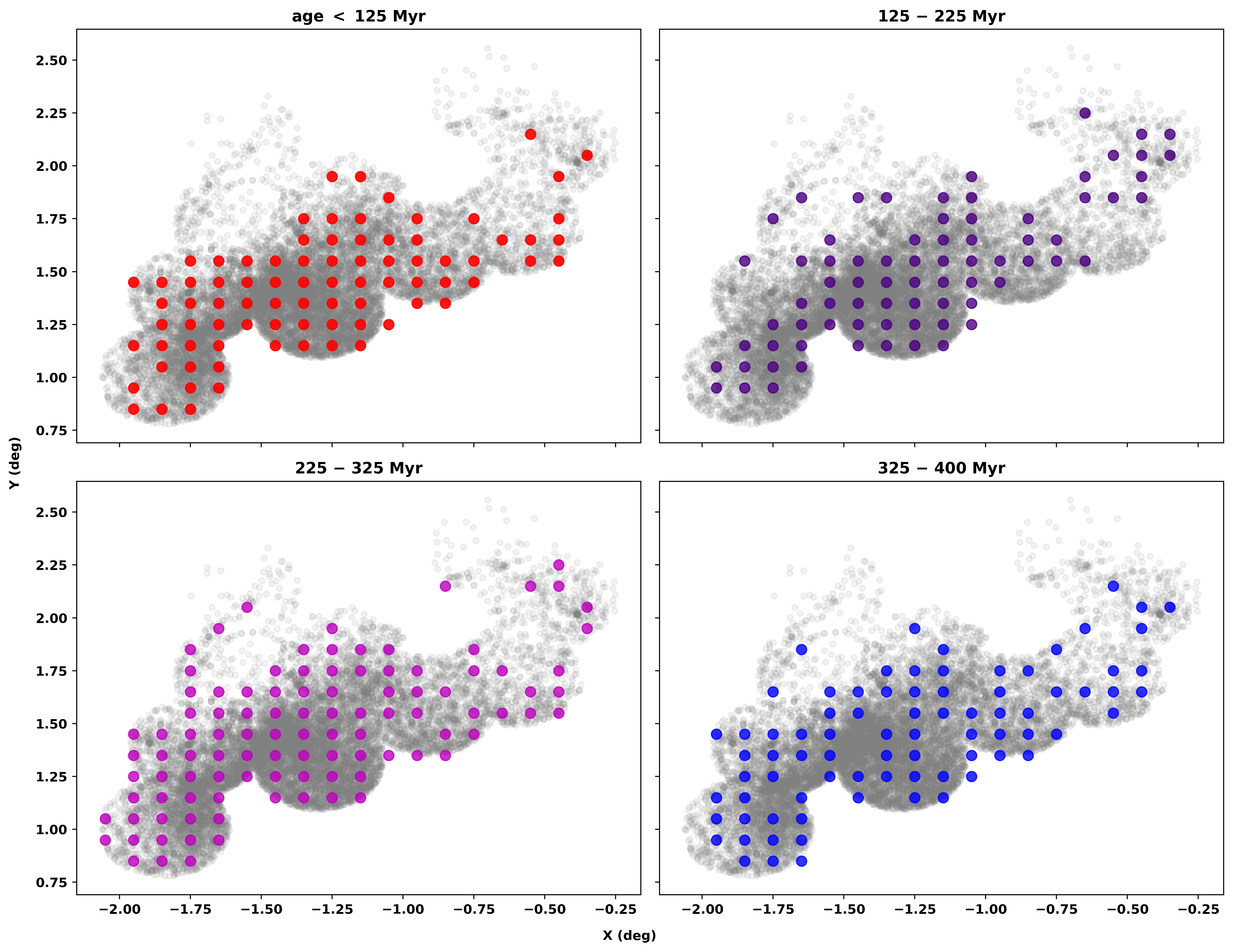}
    \caption{Spatial distribution of the age of the major star forming events of individual regions, estimated from the FUV--optical CMDs using the same parameters as mentioned in Section~\ref{section:cmd}. All stars detected in the FUV are shown as gray-coloured dots.}
    \label{fig:sf_episodic}
\end{figure*}

To check for evidence of propagating star formation, we compared the spatial distribution of bins with age groups of  $<$ 125 Myr, 125 -- 225 Myr, 225 -- 325 Myr, and 325 -- 400 Myr as shown in the Fig.~\ref{fig:sf_episodic}. There is no noticeable pattern to suggest any propagation of star formation within the SMC-Shell region in the considered age range. If we check the youngest group (age $<$ 125 Myr), we notice that the recent star formation happened in the SMC-Shell region except region towards at X $\approx$ [$-1.75$, $-1.25$] deg, Y $\approx$ [1.50, 2.25] deg including the arm-like feature and in the NE, where observations are spatially incomplete. Fig.~\ref{fig:sf_episodic} suggests that the arm-like structure is formed in the age range 125 -- 400 Myr. Instead, the arc-like structure appears to be present in all four age ranges.


\begin{figure*}
    \centering
    \begin{minipage}[t]{0.5\textwidth}
        \includegraphics[width=\textwidth,trim={0 0.1cm 0 0},clip]{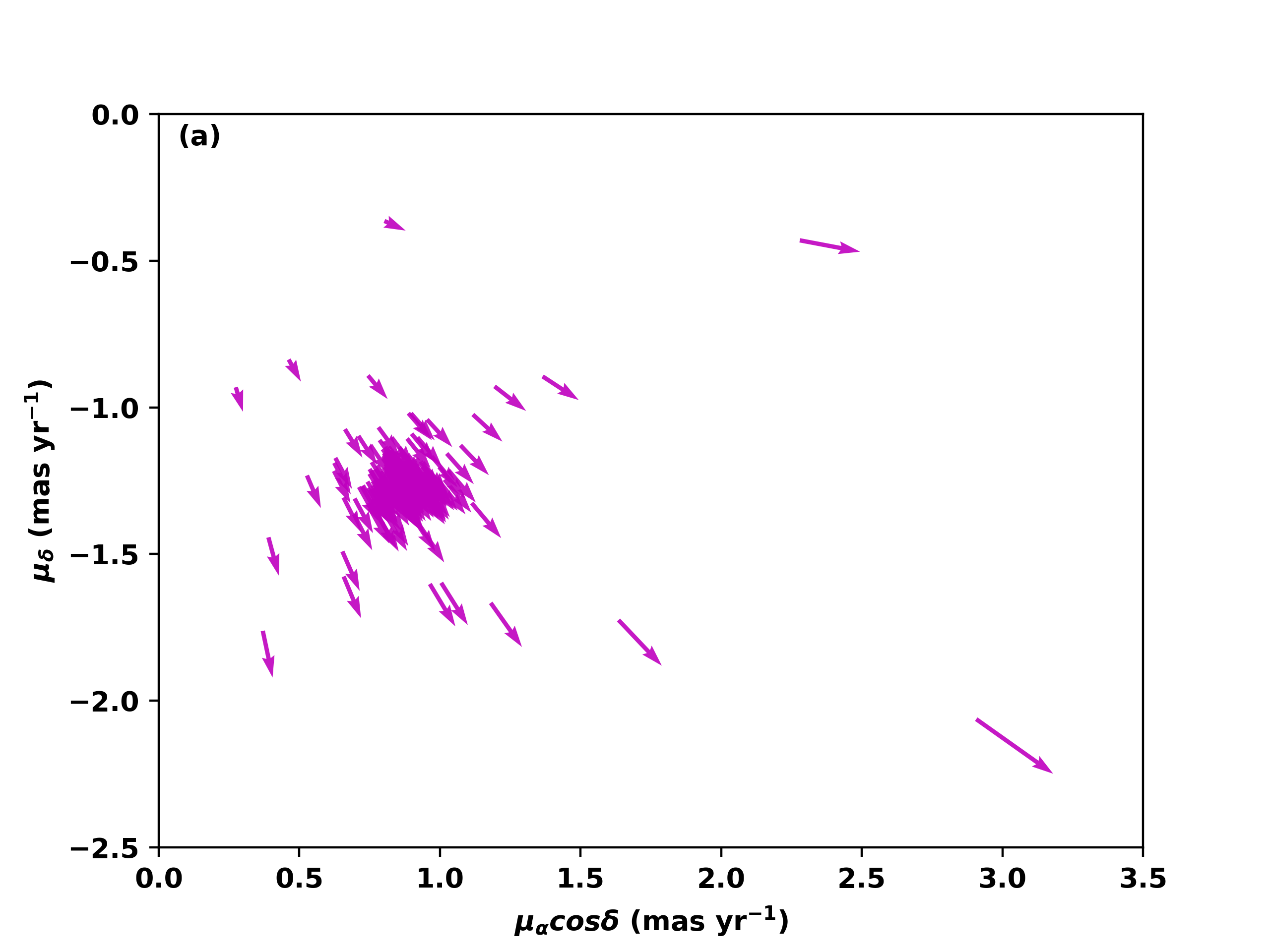}
    \end{minipage}%
    \begin{minipage}[t]{0.5\textwidth}
        \includegraphics[width=\textwidth,trim={0 0.1cm 0 0},clip]{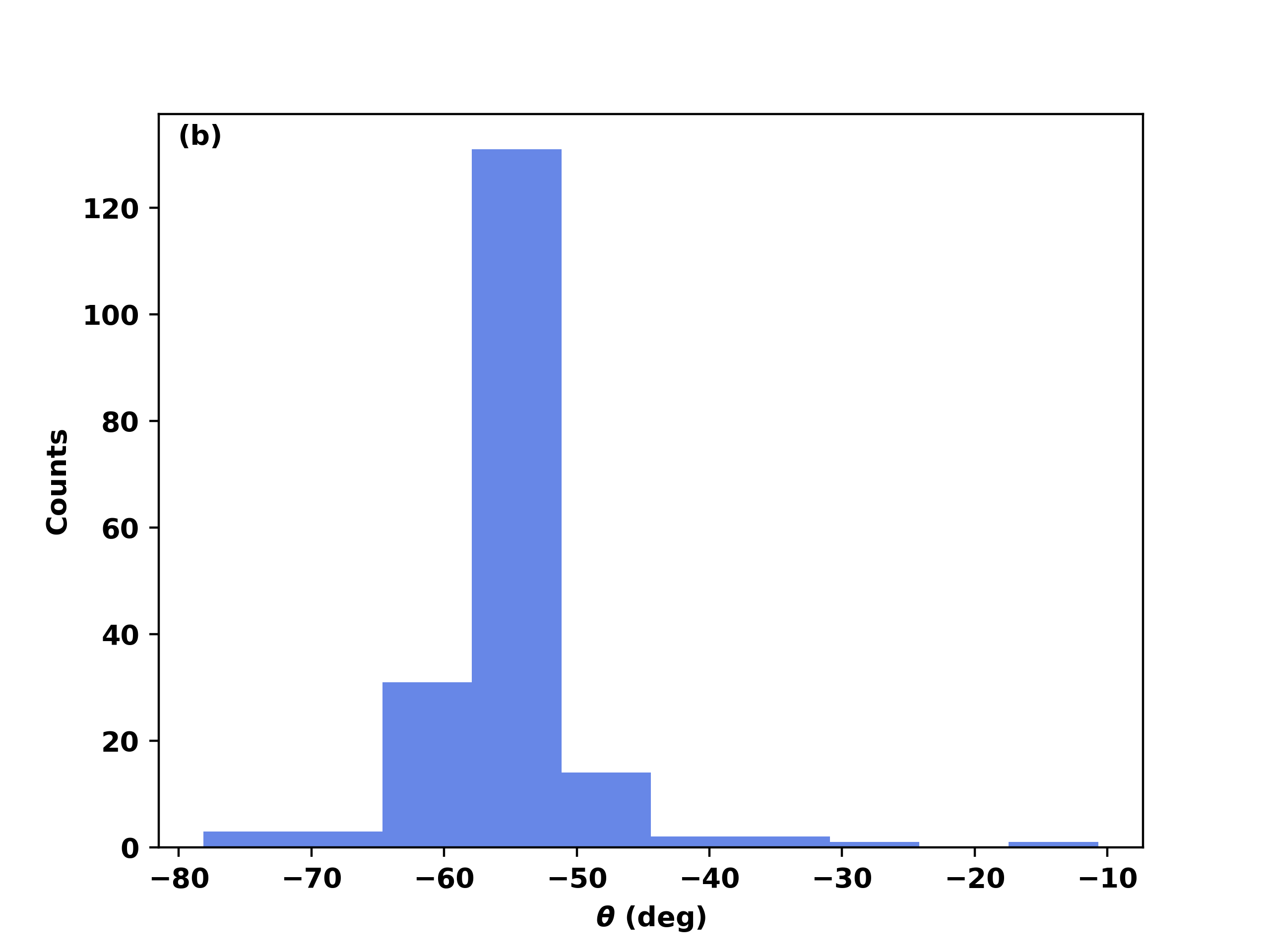}\\
    \end{minipage}
    \\
    \begin{minipage}[t]{0.5\textwidth}
        \includegraphics[width=\textwidth,trim={0 0.1cm 0 1.0cm},clip]{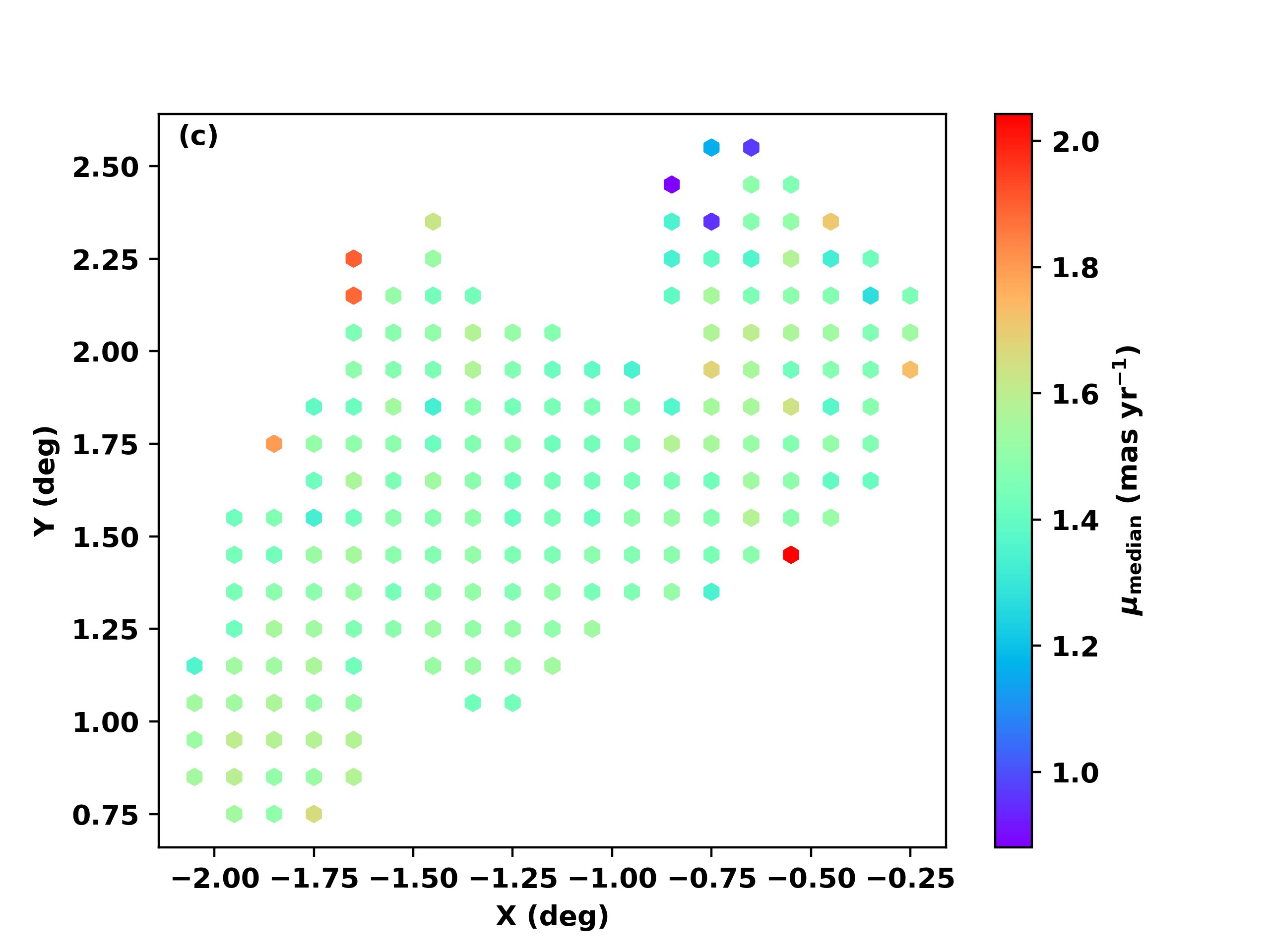}
    \end{minipage}%
    \begin{minipage}[t]{0.5\textwidth}
        \includegraphics[width=\textwidth,trim={0 0.1cm 0 1.0cm},clip]{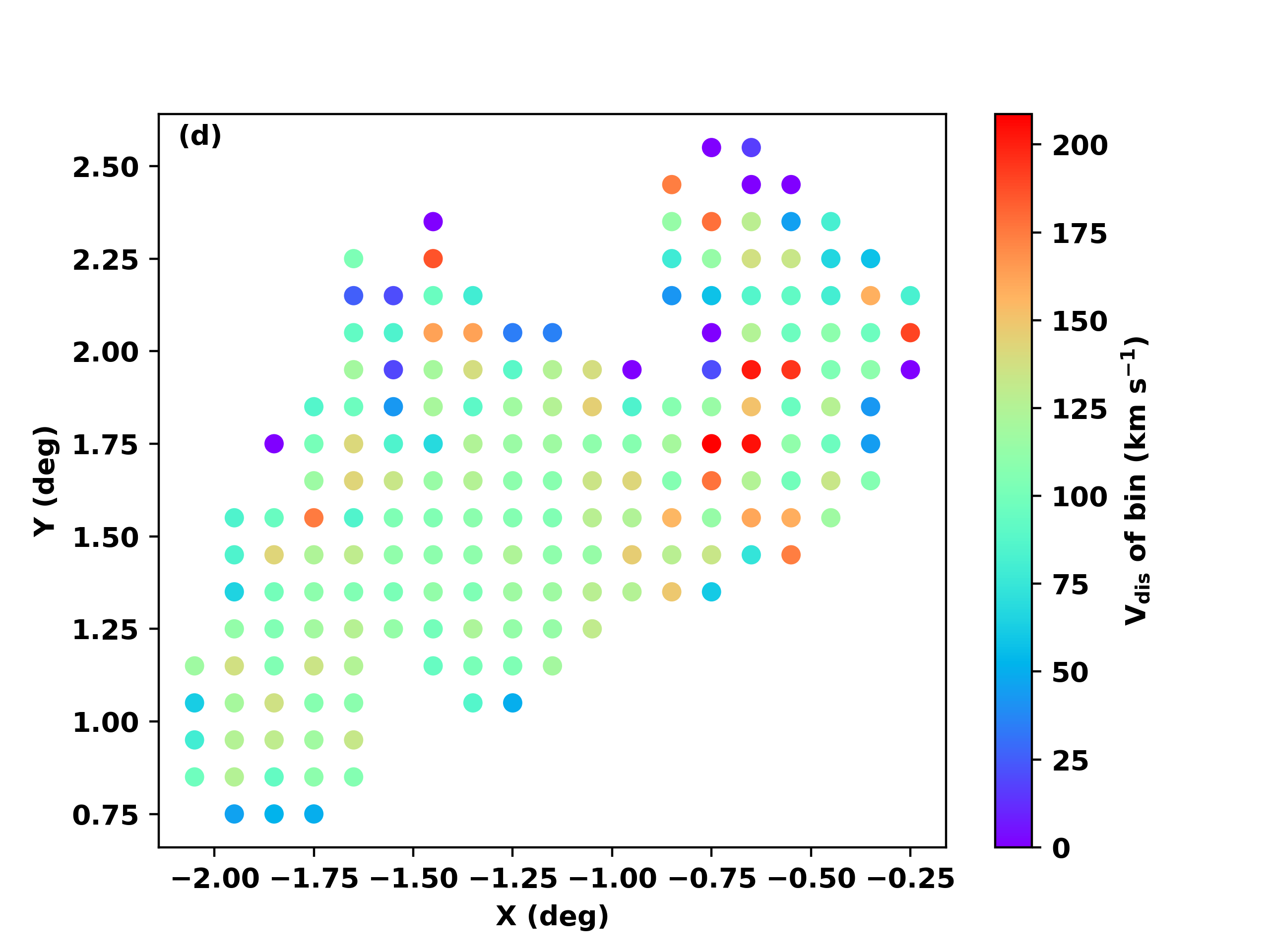}
    \end{minipage}
    \\
    \begin{minipage}[t]{0.5\textwidth}
        \includegraphics[width=\textwidth,trim={0 0.1cm 0 0.8cm},clip]{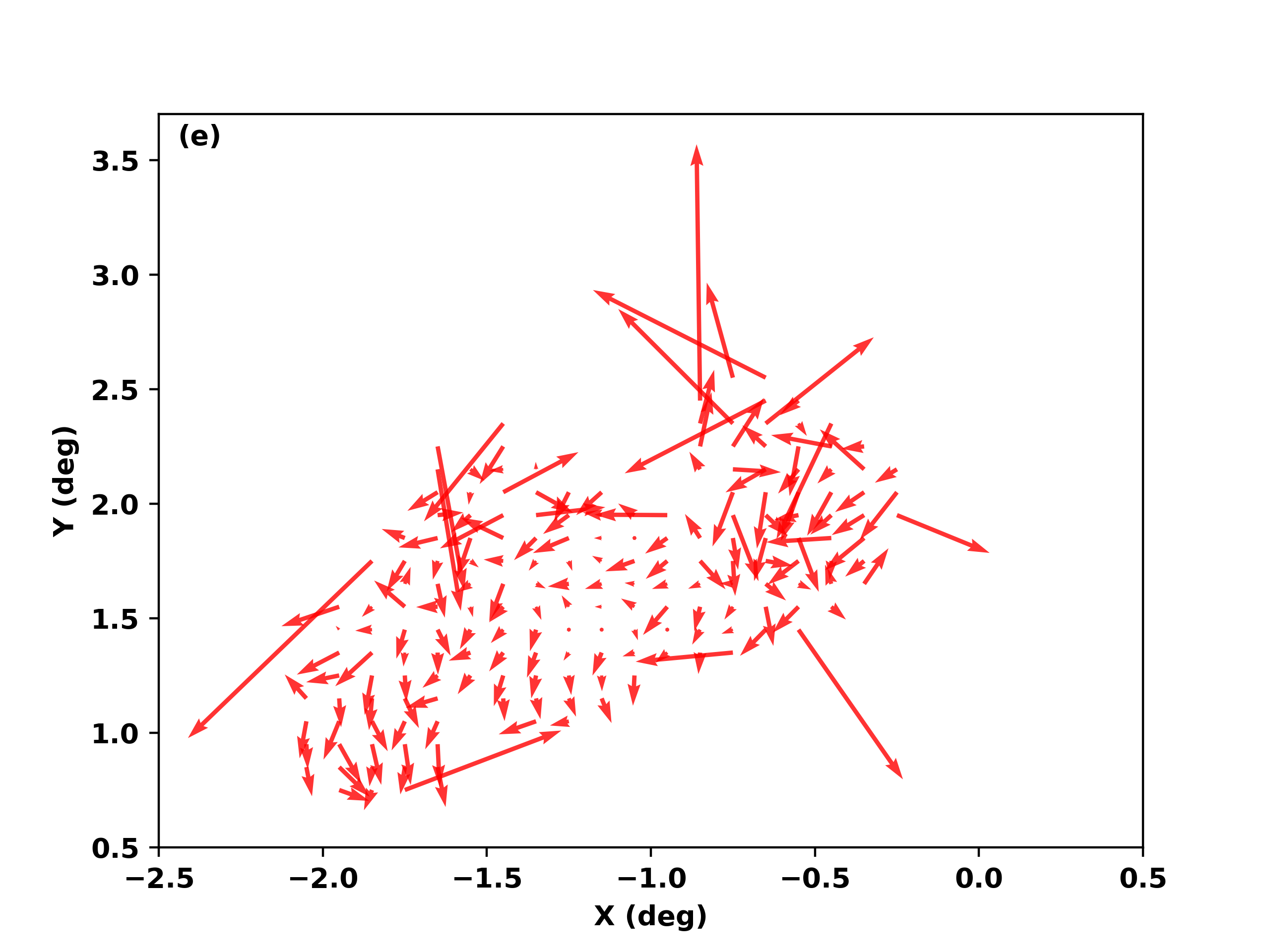}
    \end{minipage}%
    \begin{minipage}[t]{0.5\textwidth}
        \includegraphics[width=\textwidth,trim={0 0.1cm 0 0.8cm},clip]{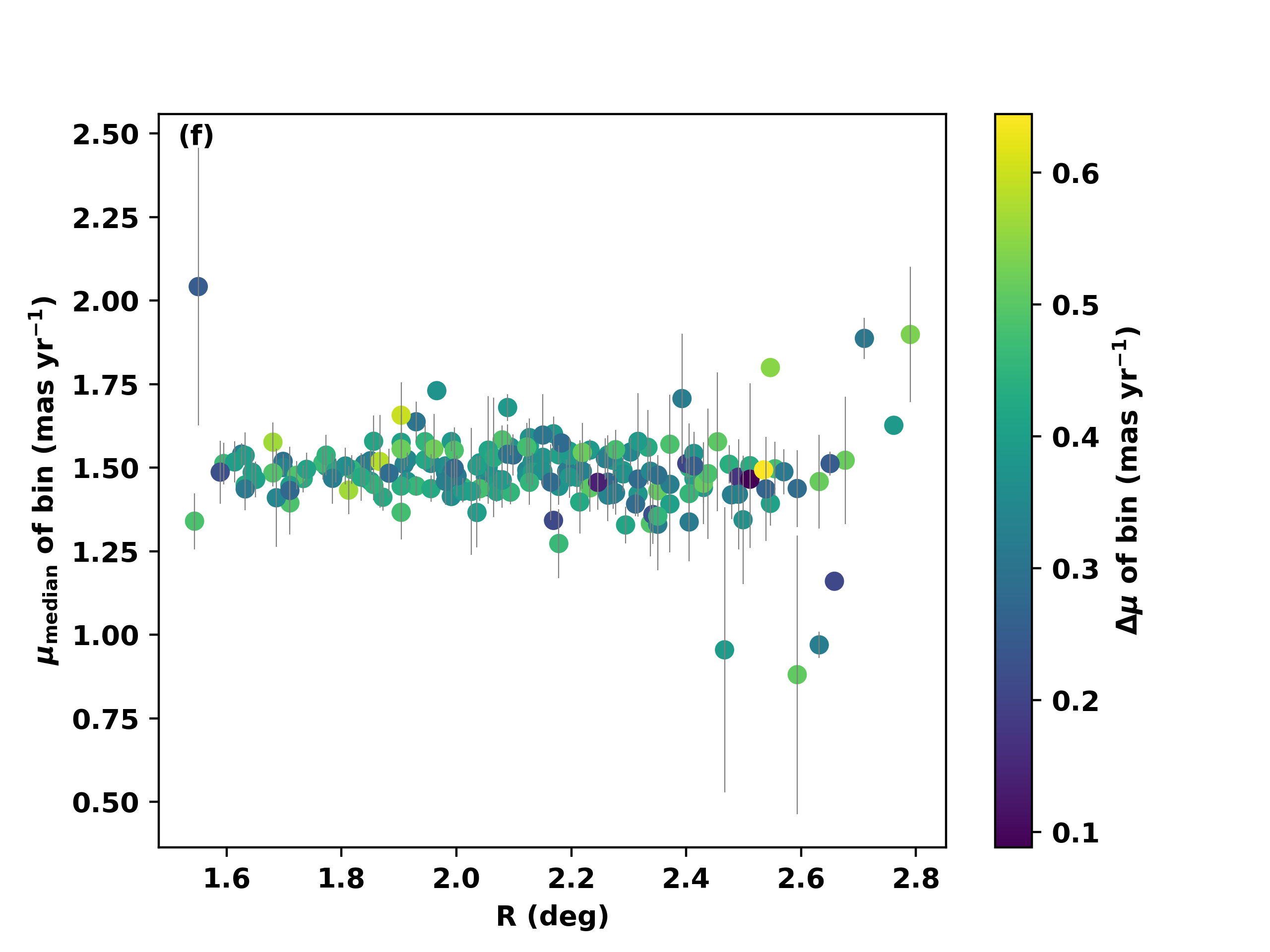}
    \end{minipage}
    \caption{(a) A plot of the median PM vector of each bin in the SMC-Shell region in the $\mu_{\alpha}\cos{\delta}$ and $\mu_{\delta}$ plane, (b) a histogram of the angle between the median PM vector and the $\mu_{\alpha}\cos{\delta}$, (c) the colour map of the median PM of each bin in X-Y plane, (d) the colour map of the velocity dispersion of each bin in X-Y plane, (e) a plot of the residual PM vector in the X-Y plane, and (f) a plot of the radial variation of the median PM of each bin with an error bar (standard error) where the colour bar represents the median value of error in PM within each bin. }
    \label{fig:proper}
\end{figure*}

To probe the kinematics of the SMC-Shell region, we created the VPD (Fig.~\ref{fig:proper}a), where the median PM vectors of each bin are shown. Most of the vectors are aligned and co-located, with some scatter. The position angle (PA) of the median PM vector (angle of median PM to $\mu_{\alpha}\cos{\delta}$) is found to be between $-$65 deg to $-$45 deg for the majority of the bins (Fig.~\ref{fig:proper}b). For the spatial distribution of PM, we transformed ($\mu_{\alpha}cos\delta$, $\mu_{\delta}$) into the Cartesian plane using the conversion equation as defined by \citet[Eq 2,][]{2021A&A...649A...1G} and estimated the median value of the PM of each bin. Fig.~\ref{fig:proper}c shows the spatial distribution of the median PM  after a 3$\sigma$ cutoff (three bins were eliminated),  which suggests that there is no significant gradient in the median PM within the SMC-Shell region. Further, the velocity dispersion of each bin was obtained following the formula shown in Eq~\ref{eq1}. The spatial map of the velocity dispersion (Fig.~\ref{fig:proper}d) reveals a range in velocity dispersion within the SMC-Shell region. Fig.~\ref{fig:proper}e represents the spatial distribution of the residual PM vector, which is calculated by subtracting the systemic PM \citep[][]{2018A&A...616A..12G} from the median value of PM of each bin (without considering internal rotation). This plot illustrates that within the SMC-Shell region, the western and outer regions are relatively more disturbed than the inner and eastern regions.\\
In order to check for any radial gradient in the PM value,  Fig.~\ref{fig:proper}f shows the median value of the PM of each bin as a function of the radial distance. The error bar for each value indicates the standard error. This plot shows that most of the bins have a similar median value of PM, which coincides with the PM value of the SMC main body \citep[][]{2018A&A...616A..12G}. Significant scatter is noticed for bins located at a radial distance more than 2.4 deg from the SMC center. We also note that bins that show large variations in the median value of PM do not necessarily have a large PM error. The deviation in the median PM, as shown in Fig.~\ref{fig:proper}f, is found to increase after 2.4 deg, whereas the inner regions between 1.6 deg to 2.4 deg are found to have a similar PM. The inner regions are, therefore, likely to be kinematically stable, with the median PM similar to the SMC main body. This supports the finding at Section~\ref{section:density} that the NE extent of the inner SMC is likely at $\sim$ 2.2 deg. 

\section{morphological structures}
\label{section:mor}
In this study, we identified visually two morphological structures from the distribution of FUV stars in the SMC-Shell region (Fig.~\ref{fig:cartesian map}). One is an arm-like structure, and the other is an arc-like structure. To compare the kinematics of these two structures with their nearby region, we define nearby control regions. The polygons for the structures make sure that the feature (arm and arc-like) falls within, and the polygons for the control sample ensure that they are sufficiently close to the structure without contamination and have a similar number of stars. In the following, we discuss these structures in detail.
\subsection{Arm-like Structure}

\label{subsec:arm}
\begin{figure*}
    \centering
    \begin{minipage}[t]{0.5\textwidth}
        \includegraphics[width=\textwidth,trim={0 0.1cm 0 0},clip]{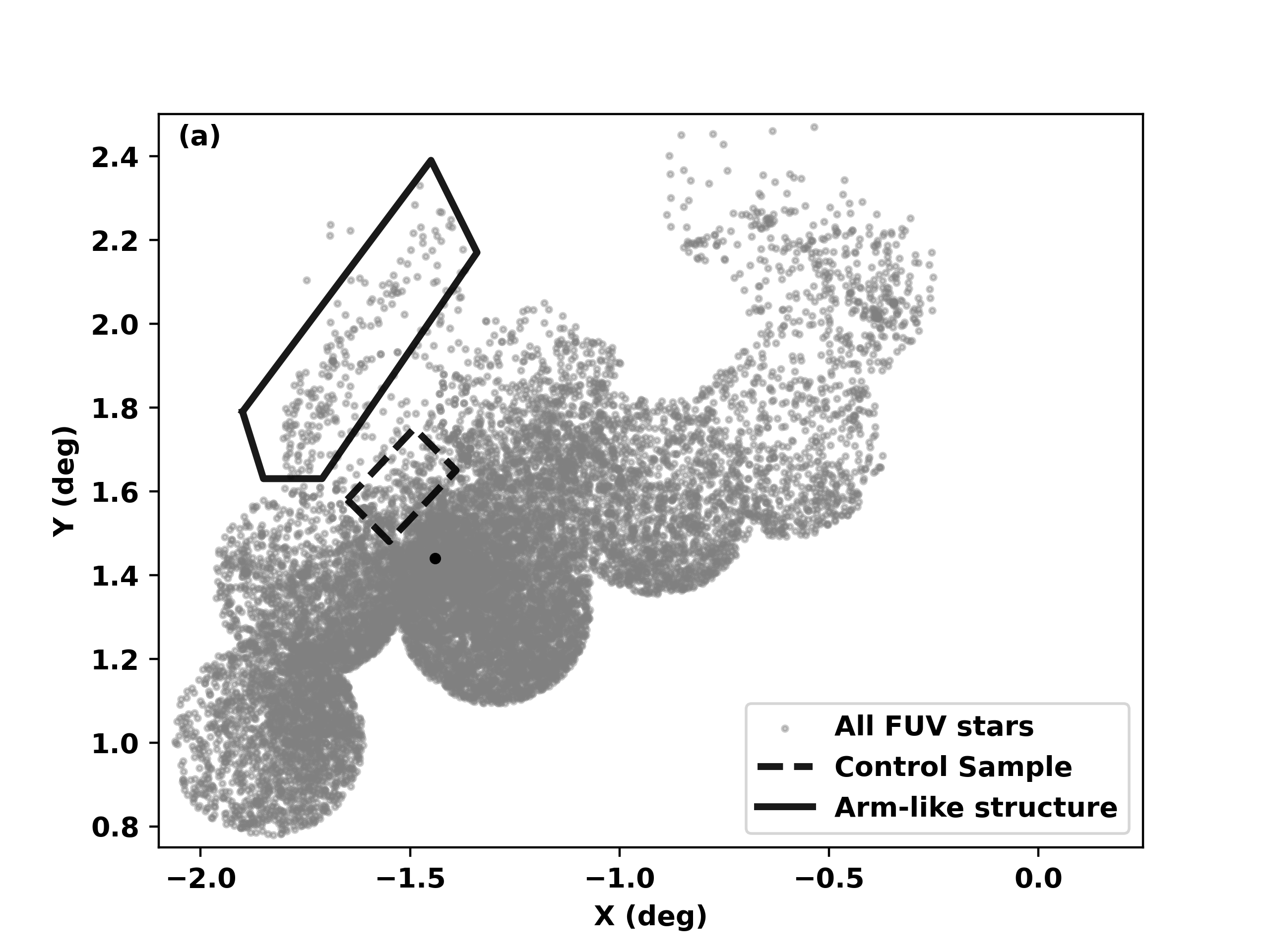}
    \end{minipage}%
    \begin{minipage}[t]{0.5\textwidth}
        \includegraphics[width=\textwidth,trim={0 0.1cm 0 0},clip]{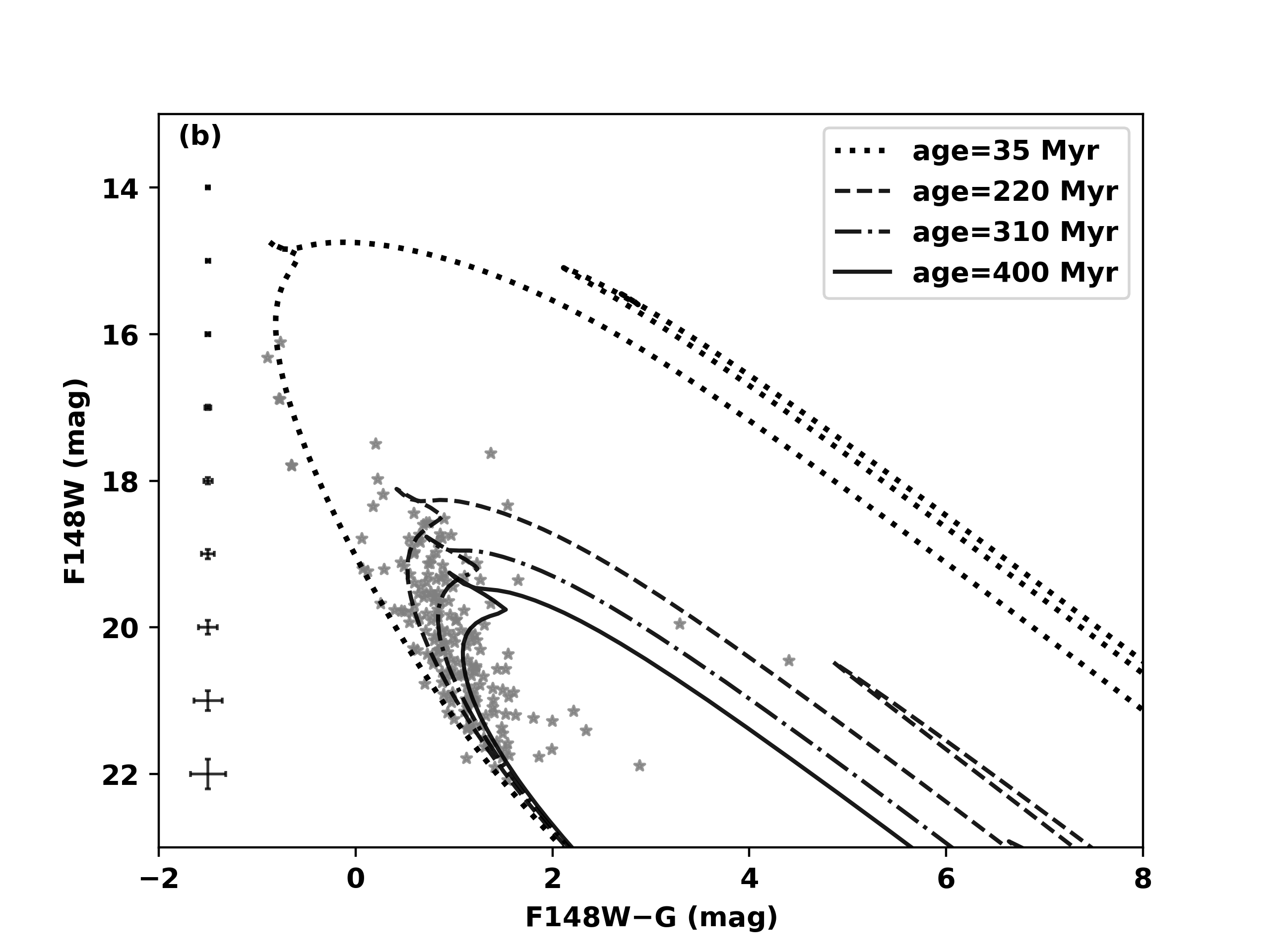}\\
    \end{minipage}
    \\
    \begin{minipage}[t]{0.5\textwidth}
        \includegraphics[width=\textwidth,trim={0 0.1cm 0 1.2cm},clip]{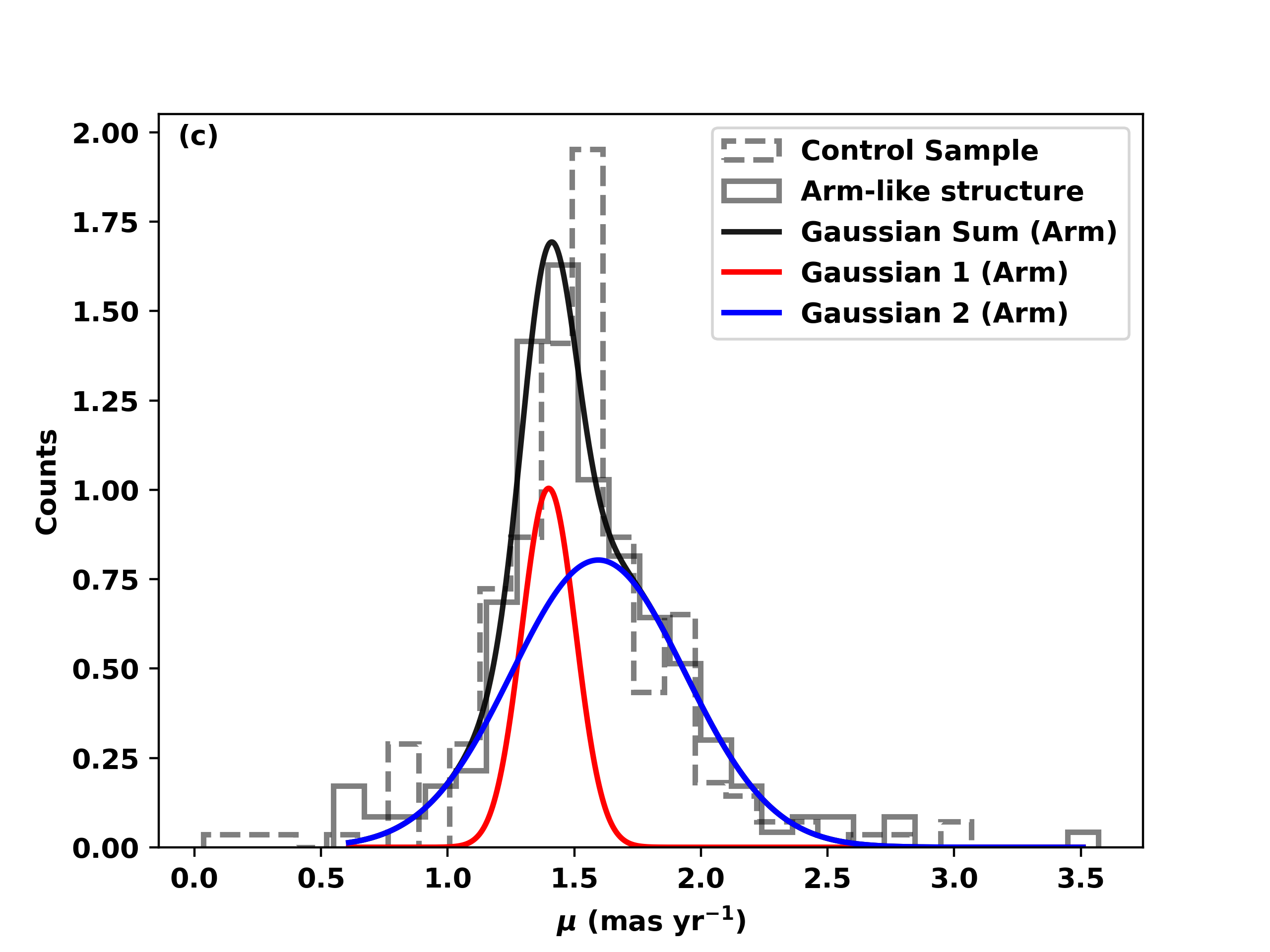}
    \end{minipage}%
    \begin{minipage}[t]{0.5\textwidth}
        \includegraphics[width=\textwidth,trim={0 0.1cm 0 1.2cm},clip]{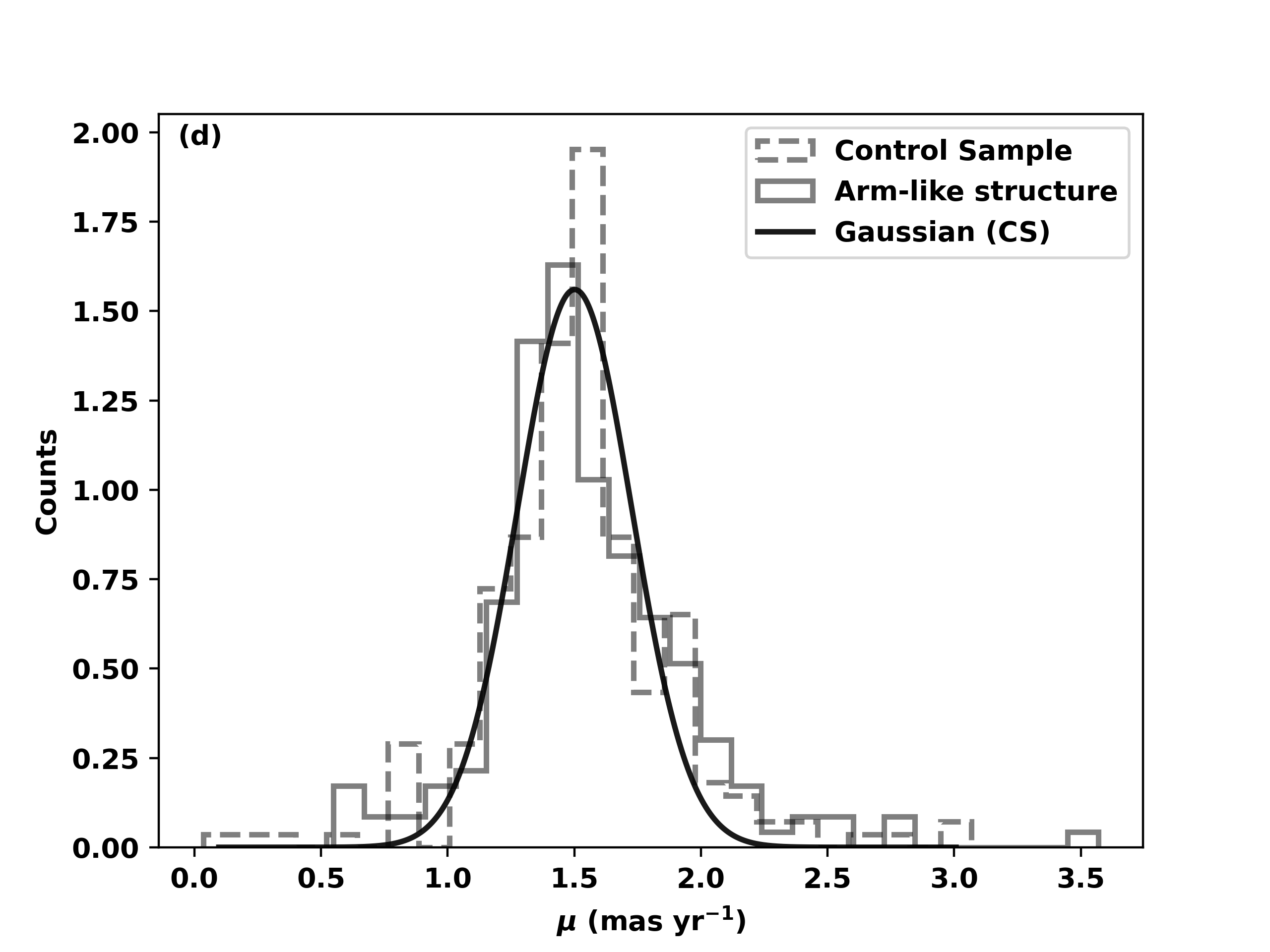}
    \end{minipage}
    \begin{minipage}[t]{0.5\textwidth}
        \includegraphics[width=\textwidth,trim={0 0.2cm 0 1.2cm},clip]{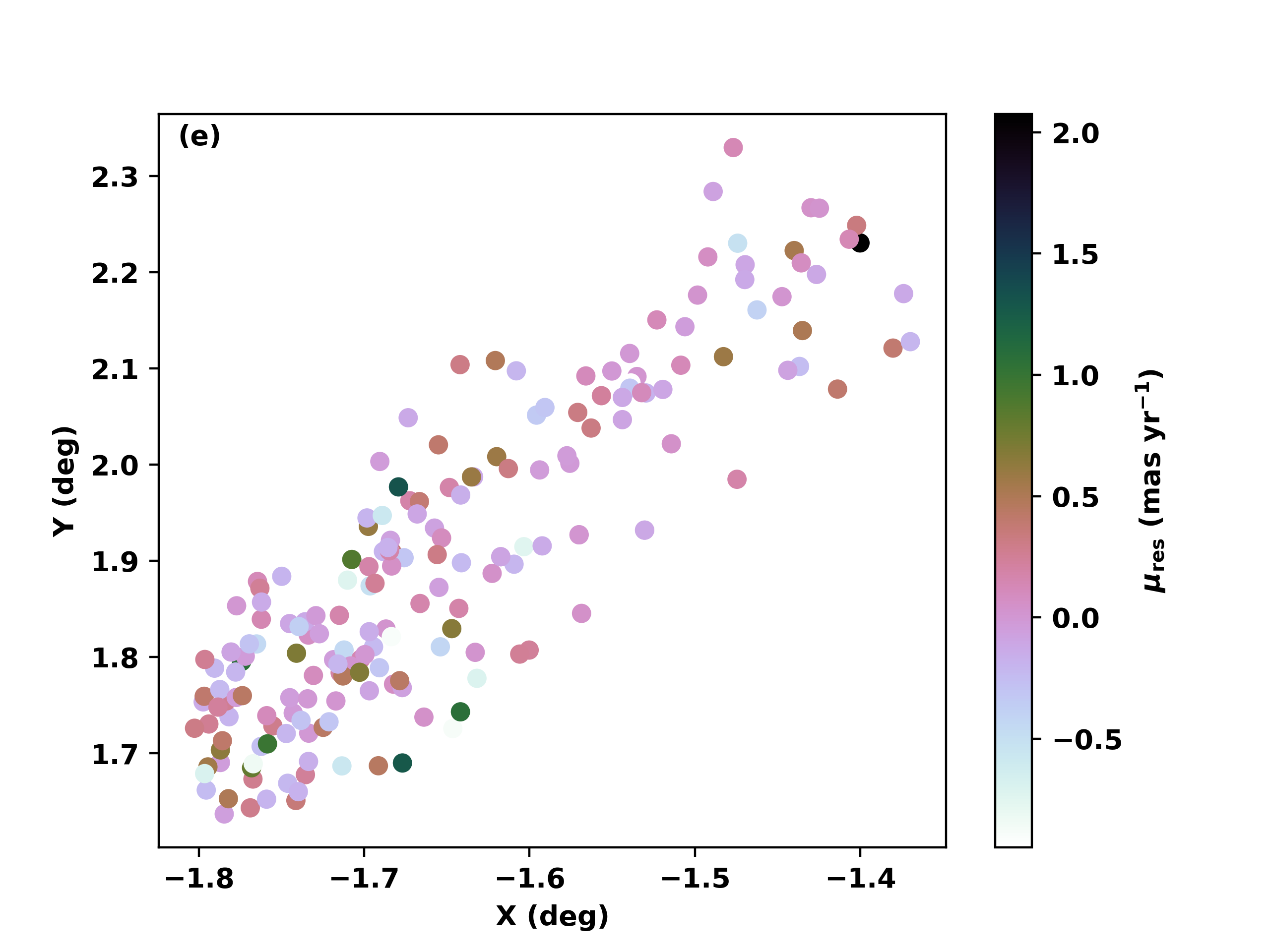}
    \end{minipage}%
    \begin{minipage}[t]{0.5\textwidth}
        \includegraphics[width=\textwidth,trim={0 0.2cm 0 1.2cm},clip]{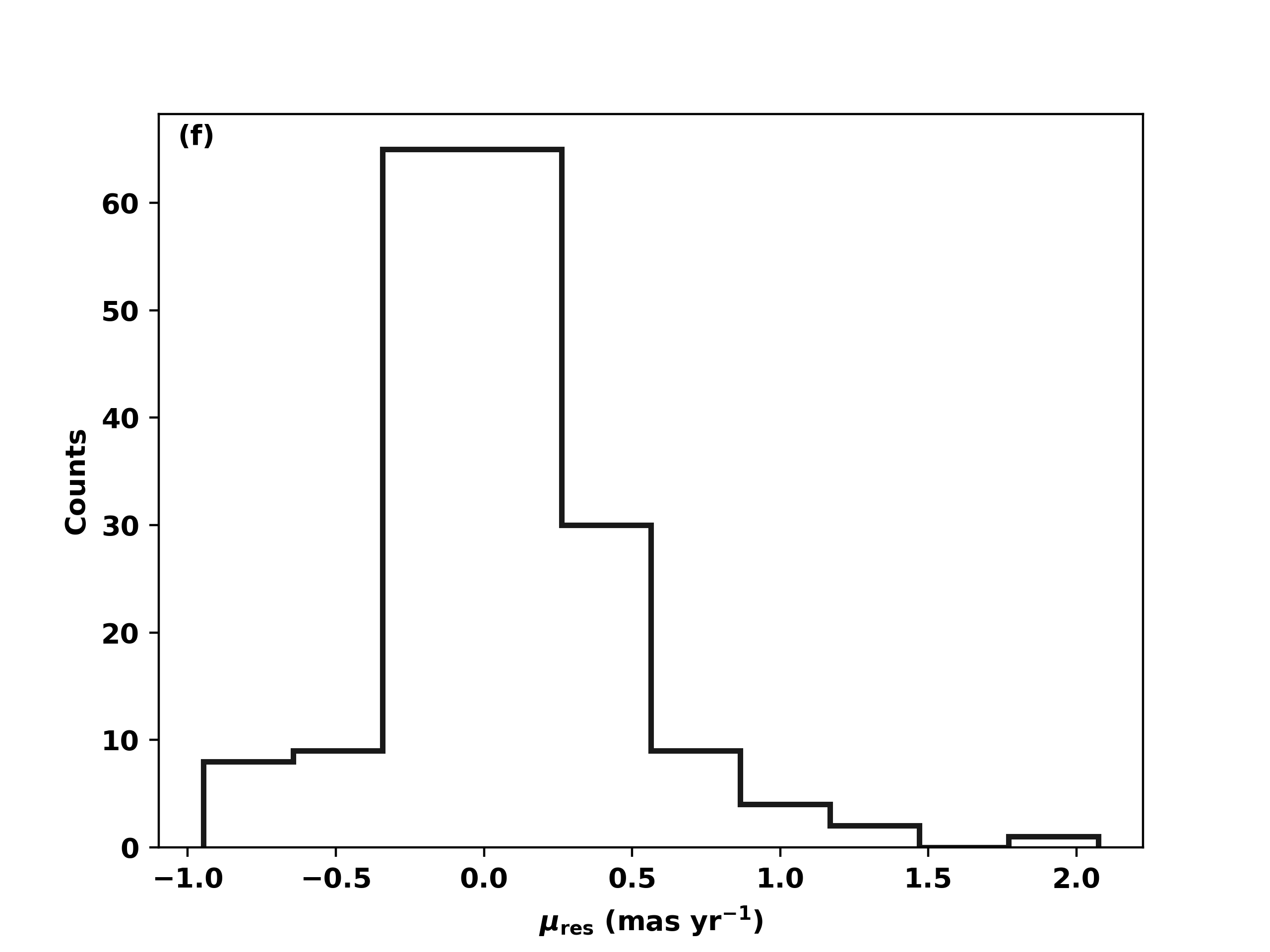}
    \end{minipage}
    \caption{Characterisation of arm-like structure: (a) Stars inside the black solid polygon are considered as belonging to the arm-like structure, and stars within the black dashed polygon are taken as a control sample for kinematic comparison, and the black dot represents the location of star cluster HW64 for a reference. (b) FUV--optical CMD of stars within the arm-like structure. (c) and (d) show Gaussian fits to the PM  histogram of the arm-like structure and control sample, respectively. (e) and (f) present the spatial distribution and histogram of the residual PM of the arm-like structure, respectively.}\label{fig:parm}
\end{figure*}

In Fig.~\ref{fig:parm}a, we show the arm-like structure enclosed within a black solid polygon and define a control sample made of stars within the black dashed polygon. To estimate the age range of stars within the arm-like structure, we have plotted the FUV--optical CMD and overlaid the Padova-PARSEC isochrones (Fig.~\ref{fig:parm}b) using the same parameters as mentioned in Section~\ref{section:cmd}. The isochrones are visually checked to match the evolutionary pattern of stars seen in the CMD. There are only a few stars with age $\sim$ 35 Myr. Thus, the arm-like structure does not have much of a contribution from stars younger than 200 Myr. This structure, therefore, shows the enhanced star formation between 220 Myr to 400 Myr ago, either continuous or episodic star formation. There may have been older stars, but here, we do not attempt to estimate older ages due to large photometric errors for stars fainter than 20 mag in FUV.

The PM distribution is fitted with a Gaussian curve to find the peak and standard deviation of the arm-like structure and the control sample (Fig.~\ref{fig:parm}c and \ref{fig:parm}d). We found that the values of peak and standard deviation of PM, ($<\mu>$, $\sigma$) of the arm-like structure and the control sample are ($<\mu>$, $\sigma$)$_{arm\_wide}$= (1.595 $\pm$ 0.024, 0.343 $\pm$ 0.017) mas yr$^{-1}$, ($<\mu>$, $\sigma$)$_{arm\_narrow}$ = (1.398 $\pm$ 0.007, 0.106 $\pm$ 0.010) mas yr$^{-1}$, and ($<\mu>$, $\sigma$)$_{cs}$ = (1.501 $\pm$ 0.019, 0.225 $\pm$ 0.019) mas yr$^{-1}$, respectively. The narrow and wide PM components of the arm-like structure suggest a two-component population, with the narrow component constituting 28\% of the sample. The peak PM values of the narrow and wide components are different, and the difference is statistically significant, suggesting the presence of a population within the arm-like structure with reduced dispersion in PM.
However, the peak value of the control sample PM distribution is similar to the peak of the wide component of the arm-like structure. This indicates that 72\% of the arm-like structure and the control sample population are kinematically indistinguishable. To estimate the residual PM, we subtracted the median value of PM of the arm-like structure from the PM of each star. Here, the internal rotation is not taken into account. The spatial distribution of residual PM (Fig.~\ref{fig:parm}e) shows that most stars have residual PM close to zero. Fig.~\ref{fig:parm}f shows the residual to peak near zero PM. We infer that the arm-like structure is an over density formed between 220 -- 400 Myr ago with kinematics similar to the underlying part of the SMC.

\subsection{Arc-like structure}

\begin{figure*}
    \centering
    \begin{minipage}[t]{0.5\textwidth}
        \includegraphics[width=\textwidth,trim={0 0.2cm 0 1.0cm},clip]{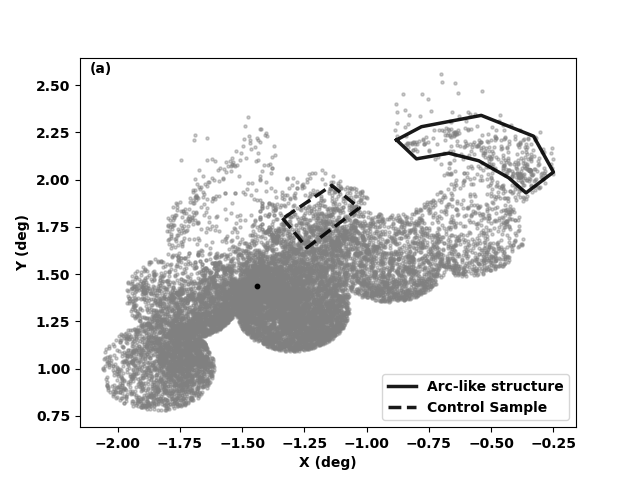}
    \end{minipage}%
    \begin{minipage}[t]{0.5\textwidth}
        \includegraphics[width=\textwidth,trim={0 0.2cm 0 1.0cm},clip]{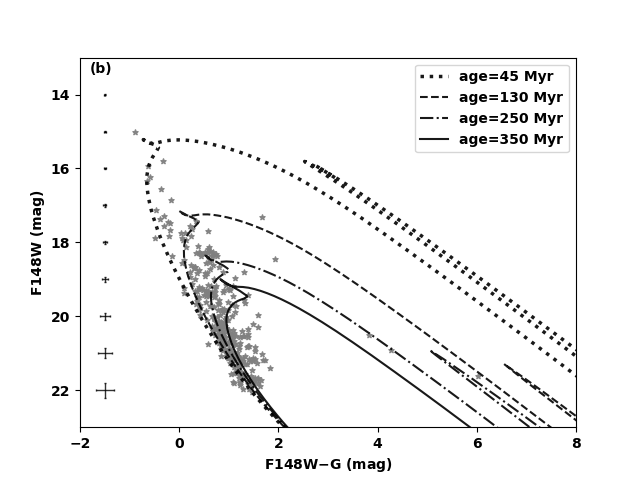}\\
    \end{minipage}
    \\
    \begin{minipage}[t]{0.5\textwidth}
        \includegraphics[width=\textwidth,trim={0 0.1cm 0 1.0cm},clip]{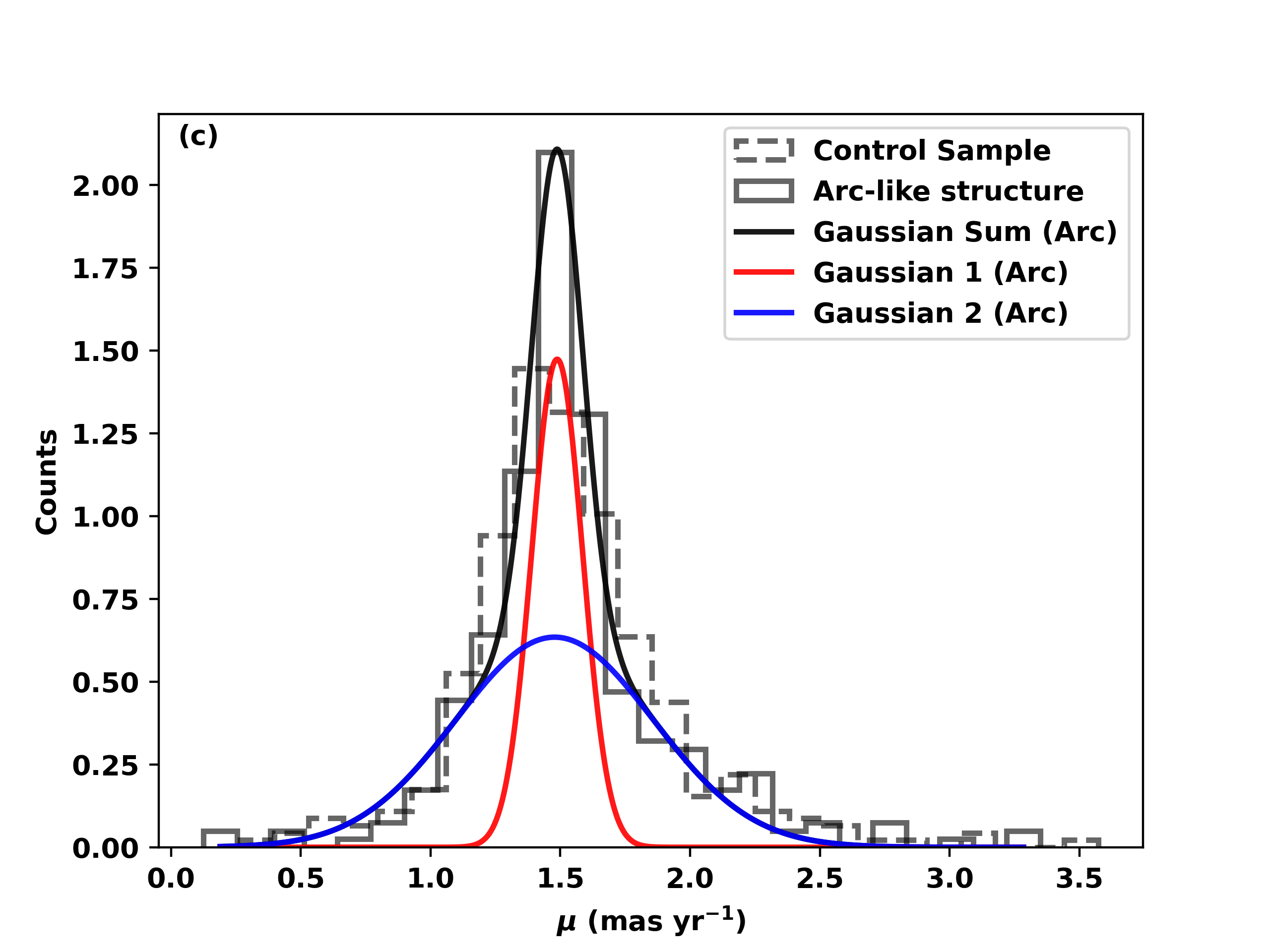}
    \end{minipage}%
    \begin{minipage}[t]{0.5\textwidth}
        \includegraphics[width=\textwidth,trim={0 0.1cm 0 1.0cm},clip]{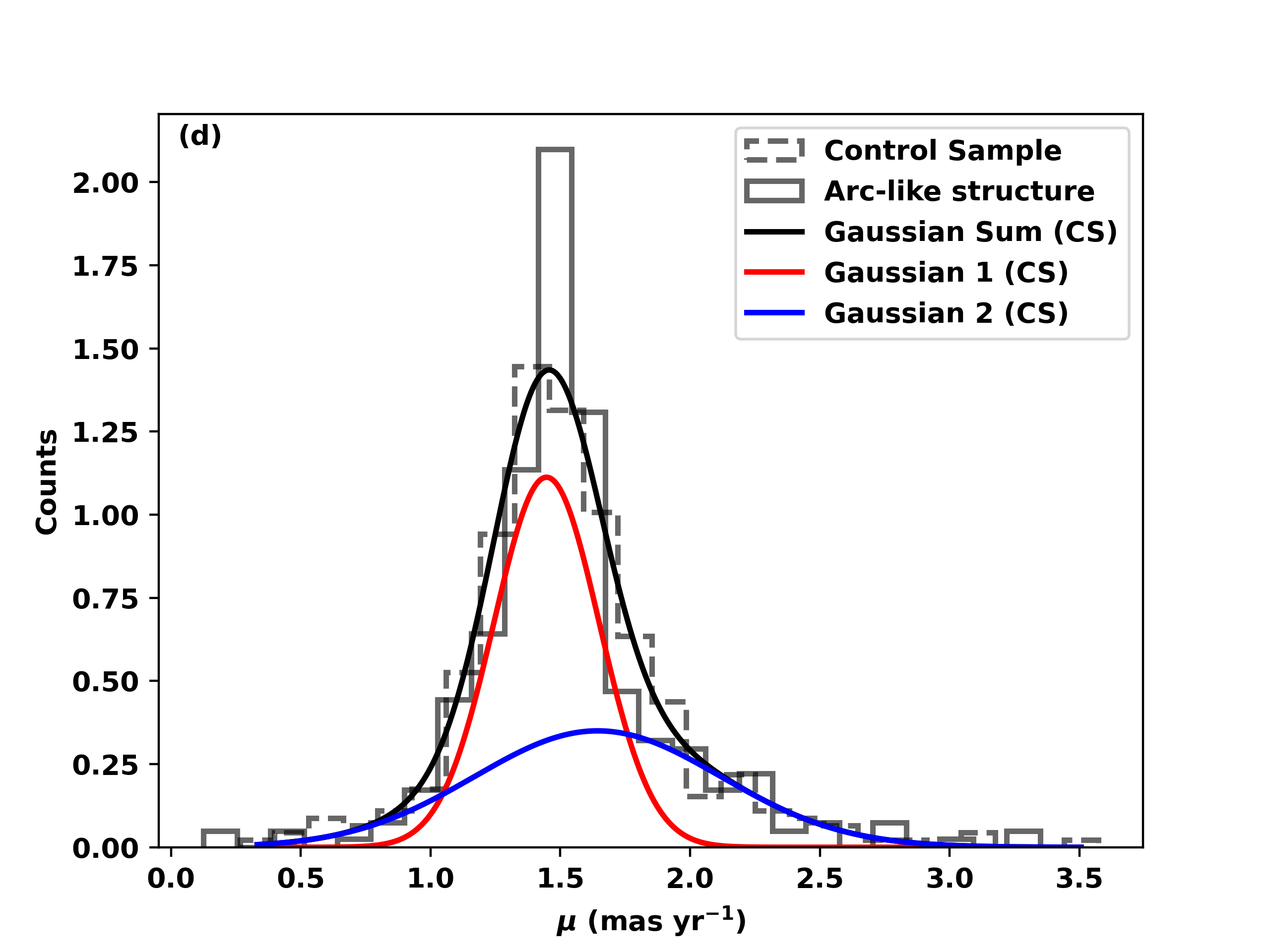}
    \end{minipage}
    \begin{minipage}[t]{0.5\textwidth}
        \includegraphics[width=\textwidth,trim={0 0.20cm 0 1.0cm},clip]{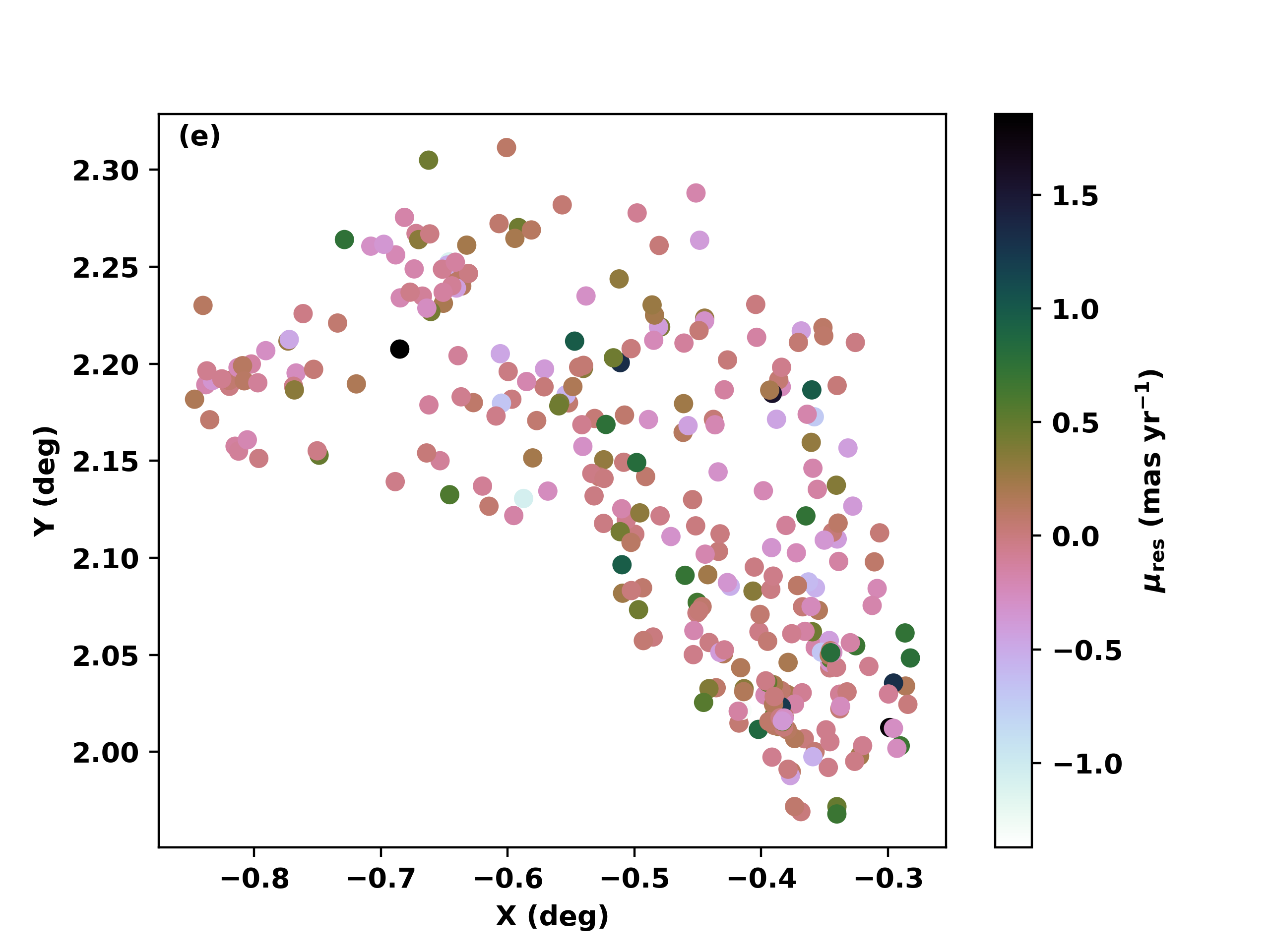}
    \end{minipage}%
    \begin{minipage}[t]{0.5\textwidth}
        \includegraphics[width=\textwidth,trim={0 0.1cm 0 1.0cm},clip]{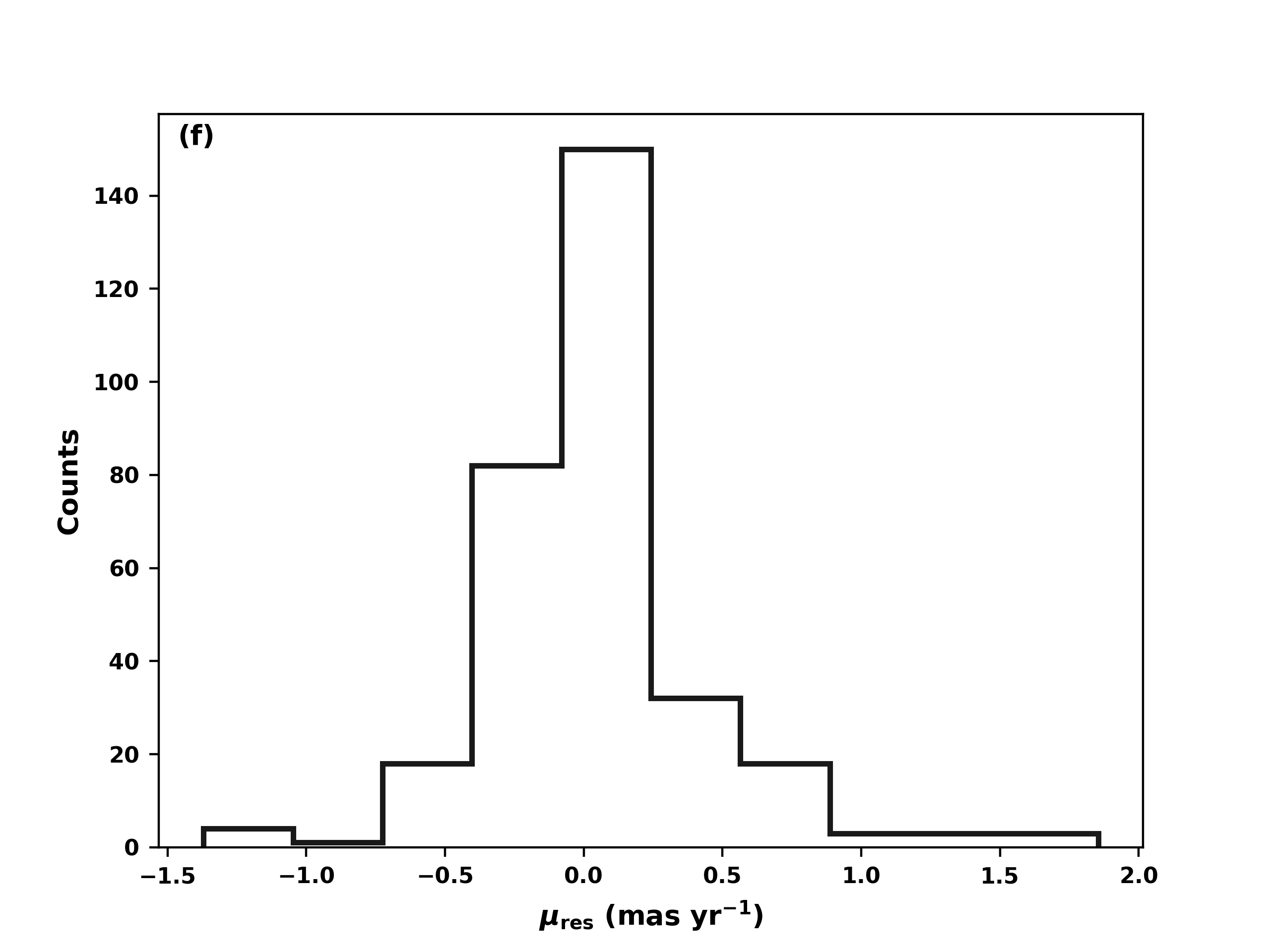}
    \end{minipage}
    \caption{Characterisation of arc-like structure: (a) Stars inside the black solid polygon are considered as belonging to the arc-like structure, and stars within the black dashed polygon are taken as a control sample for kinematic comparison, and the black dot represents the location of star cluster HW64 for a reference. (b) FUV--optical CMD of stars within the arc-like structure, (c) and (d) show Gaussian fit (multiple Gaussian curves) to the PM histogram of the arc-like structure and the control sample, respectively. (e) and (f) present the spatial distribution and histogram of the residual PM of the arc-like structure, respectively.}\label{fig:Arc-like structure}
\end{figure*}
\citet[][]{2019A&A...631A..98M} detected this separated small arc, situated ~\ang{;30;} west of the globular cluster NGC~362. In Fig.~\ref{fig:Arc-like structure}a, we have considered the arc-like structure within the black solid polygon and the nearby region within the black dashed polygon as a control sample to study the kinematics. To estimate the age, the FUV--optical CMD (Fig.~\ref{fig:Arc-like structure}b) is overlaid with the Padova-PARSEC isochrones taking the same parameters as mentioned in Section~\ref{section:cmd}. We detect a small number of stars with age $\sim$ 45 Myr.
From Fig.~\ref{fig:Arc-like structure}b, the arc-like structure appears to have a significant contribution from stars aged between 130 to 350 Myr, and the contribution from the FUV stars younger than 130 Myr is very small. This region also shows episodes of star formation like the arm-like structure (Fig.~\ref{fig:parm}b). This region may be formed in several episodes of the star formation, including the last LMC-SMC interaction.

Using the PM values of the FUV stars of the arc-like structure and its control sample, the PM distributions were created and fitted with a Gaussian curve to find the peak and standard deviation of PM of these two regions (Fig.~\ref{fig:Arc-like structure}c and \ref{fig:Arc-like structure}d). We found that the peak and the standard deviation values of PM of arc-like structure and its control sample are ($<\mu>$, $\sigma$)$_{\mathrm{arc\_narrow}}$= (1.487 $\pm$ 0.005, 0.097 $\pm$ 0.006) mas yr$^{-1}$ (fraction = 0.37), ($<\mu>$, $\sigma$)$_{\mathrm{arc\_wide}}$= (1.477 $\pm$ 0.025, 0.380 $\pm$ 0.035) mas yr$^{-1}$; and ($<\mu>$, $\sigma$)$_{\mathrm{cs\_narrow}}$= (1.446 $\pm$ 0.009, 0.203 $\pm$ 0.014) mas yr$^{-1}$ (fraction = 0.56), ($<\mu>$, $\sigma$)$_{\mathrm{cs\_wide}}$= (1.644 $\pm$ 0.070, 0.475 $\pm$ 0.054) mas yr$^{-1}$. The narrow and wide components of the arc region have similar values of PM with respect to that of the narrow component of the control sample. We note that the control sample has a broad component with significantly different PM values. It appears that there is a stellar population with broad distribution of PM in this region, that is not found in the control region of the arm-like feature, though these are located nearby.

Two components of the arc-like structure have similar peak values of the PM with a different standard deviation. This leads to the conclusion that the arc-like structure may contain two sub-populations. However, the narrow component of the control sample shows a motion similar to that of the arc-like structure, indicating that the arc-like structure is not kinematically distinguishable from its control sample. To measure the residual PM of the arc-like structure, we subtracted the median PM of the arc-like structure from the PM of individual stars within this structure. Here also, the internal rotation is not taken into account. The spatial distribution of residual PM (Fig.~\ref{fig:Arc-like structure}e) shows that most of the stars have the same PM as their median, as the residual is almost zero (see Fig.~\ref{fig:Arc-like structure}f).

We noted from the FUV--optical CMD (Fig.~\ref{fig:Arc-like structure}b) that, though the arc-like structure has stars with age ranging from 45 to 350 Myr, most of the stars have ages in the range 130 -- 350 Myr. If the arc is formed from holes created due to material driven away by OB stars or supernovae, the age range of stars formed from the scooped-up material in the shell of such an arc will not be so large \citep[][]{2002ARep...46..791E}. On the other hand, it may be possible that different parts of the arc underwent collapse over a period of time. However, Fig.~\ref{fig:sf_episodic} suggests that there is no localized age-specific star formation in the arc-like structure. Therefore, this structure is probably a random over density.

\section{Discussion}
\label{section:diss}
This paper presents a comprehensive study of the young population residing in the SMC-Shell region using FUV observations obtained from the UVIT and optical data from Gaia. This study presents the FUV--optical catalog (Table~\ref{tab:catalog}) and FUV--optical CMD (Fig.~\ref{fig:cmd}a) of the SMC-Shell region. From the KDE map (Fig.~\ref{fig:cartesian map}) of the SMC-Shell region, we identify the arm-like and the arc-like structures, which have previously been detected within the SMC-Shell region by \cite{2019A&A...631A..98M}.

The density map (Fig.~\ref{fig:cartesian map} and Fig.~\ref{fig:den}a) represents a clumpy distribution of the young stellar population, unlike the older population in the SMC-Shell region 
\citep[][]{2000ApJ...534L..53Z,2007AJ....133.2037N}. The FUV number density reduces radially outward \citep[Fig. 4,][]{2019A&A...631A..98M}. \citet{2009ApJ...705.1260N} found a similar surface density of the young population of the SMC Wing. Fig.~\ref{fig:den}b suggests that we are able to trace the extent of the inner SMC in the NE direction at  $\sim$ 2.2 deg. Also, Fig.~\ref{fig:den}a reveals that the arm-like structure lies outside of the  R $=$ 2.2 deg radius, whereas the arc-like structure starts inside the R $=$ 2.2 deg radius and extends beyond. We suggest that the arm-like and arc-like structures could be equivalent to the faint extended UV features found in external galaxies \citep{2007ApJS..173..538T}.

Our study indicates that the kinematics of the young stars in the NE SMC is similar to the SMC main body. The median PM spatial map (Fig.~\ref{fig:proper}c) suggests that the NE part of the SMC is likely to be undisturbed (with the limits of this study; see Appendix~\ref{append: appendix}) and, therefore, unlikely to have experienced significant tidal perturbation in comparison to the wing region, which is extensively disturbed \citep[][]{2021MNRAS.502.2859N}. A relatively large scatter in median PM is visible in Fig.~\ref{fig:proper}f after 2.4 deg, which implies that the NE extent of the inner SMC is close to the photometric inner edge of  $\sim$ 2.2 deg and the disk beyond it is tidally disturbed. \citet{2020MNRAS.495...98..De..Leo..tidal..radius} found the tidal radius of the SMC to be at $\sim$ 2 kpc, while analyzing the spectra of $\sim$ 3000 SMC RGB stars. \citet{2022MNRAS.512.4334Das..D..tidal..radius} studied the star clusters of the SMC West Halo and found the signature of tidal disruption beyond $\sim$ 2 kpc. Our result is therefore in good agreement with the above findings. The estimate of a radial distance of $\sim$ 2.2 deg as the stable kinematic extent of the SMC in the NE region will be an important input to the LMC-SMC interaction models. More studies of the outer SMC at larger radii than studied here will shed light on the outer SMC properties.
 
 We derive that the FUV bright ($<$ 150 Myr) and sub-giant stars (150 -- 300 Myr) display comparable kinematics. The transverse velocity dispersions obtained from our work for the above ages are similar to the radial velocity dispersion estimated by \citet{2023arXiv230414368E} for the main body of the SMC. These indicate that the kinematics of the young stars in the SMC-Shell region are not altered by the latest LMC-SMC interaction.

The recent work by \citet{2023sakowska} on the SMC-Shell region revealed that it experienced increased star formation within the last few billion years. Our results (Fig.~\ref{fig:cmd} and Fig.~\ref{fig:hst_age}) are in agreement with \citet[their Fig. 3]{2023sakowska}, which highlight a significant episode of recent star formation approximately between 150 and 650 Myr ago. \citet{2022MNRAS.509.3462P} detected a peak of young cluster formation in the NE SMC-Shell region at 30 -- 200 Myr, where we detected a peak of star formation in the same age range. By considering nine young star clusters \citet{2019A&A...631A..98M} estimated the age of the SMC-Shell region as $\sim$ 150 Myr, a value within our estimated age range of star formation. The detected episode of star formation at 240 -- 280 Myr is probably linked to the last LMC-SMC interaction \citep[]{1985PASA....6..104M,2012MNRAS.421.2109B,2022ApJ...927..153C} happened at 150 -- 300 Myr.

The spatial age map of the SMC-Shell reveals that star formation in this region occurred episodically, lacking any distinct pattern of propagation in star formation, with the younger stars being formed slightly inward. The study conducted by \cite{2016RAA....16...61J} using classical Cepheids aligns with our findings, as both studies identify that the majority of star formation happened at $\sim$ 200 -- 300 Myr. The arm-like feature is likely formed between 220 -- 400 Myr, whereas the arc-like feature is formed between 130 -- 350 Myr. The arc-like structure is probably younger than the arm-like structure. The study by \cite{2019A&A...631A..98M} reported that the formation of the SMC-Shell region could be attributed to factors other than tidal effects. As this region does not show any evidence of tidal disturbance, we also rule out the tidal origin of the arm-like and arc-like features. The arm-like feature is likely to be an over density created due to multiple star formation events. We rule out any connection to the formation of the arc-like structure with star formation around shells/holes created by massive stars or supernovae.\\
The youngest episode of star formation, which is relatively more confined to the inner regions, is likely to be triggered by a mechanism other than the interaction with the LMC. \citet{2011A&A...535A.115I} found a shift of the younger population in the 40 -- 200 Myr towards the NE of the SMC. They also noted that the direction of the line connecting the LMC to the MW is the same as that connecting the SMC to the MW. Compression of the HI gas and star formation as found in the NE of the LMC \citep{2015ApJ...815...77S} due to the passage of the LMC in the MW halo, is a process that may also be happening in the NE of the SMC owing to the motion of the LMC-SMC system in the halo of the MW. \citet{2018ApJ...864...55Z} suggested that the recent pericentric passage of the MCs around the MW happened around 50 Myr ago. We suggest that the youngest episode of star formation in the SMC-Shell region at 40 -- 80 Myr is probably due to the pericentric passage of the SMC around the MW.

\section{Summary and Conclusion}
In this study, we present a far-UV map and analysis of the SMC-Shell region in the NE outskirts of the SMC. A summary of the results and conclusion are given below:
\begin{enumerate}
    \item We present the first far-UV map of the northeastern SMC, based on the images from the UVIT/AstroSat. The UV catalog from this study is combined with the Gaia Early Data Release 3 (EDR3) data set to produce an FUV--optical catalog with kinematic data. We present the photometric as well as kinematic study of young stars in this region.
    \item The ages estimated from the FUV--optical and optical CMDs reveal that this region is populated with a few stars as young as 7 Myr. We find that the majority of star formation happened between 60 -- 300 Myr ago.
    \item The KDE spatial distribution of the FUV stars traces two distinct structures that were previously suggested: an arm-like structure and an arc-like structure. 
    The distribution of the FUV stars shows an outward radial gradient within the SMC-Shell region, and we suggest the extent of the inner SMC to be at a radial distance of $\sim$ 2.2 deg. The arm-like structure is located outside this boundary, while the arc-like structure extends from the inner SMC to the outer regions.
    \item  The PM, as well as the velocity dispersion of young FUV stars (computed from the PM dispersion), are found to be similar to that of the SMC main body. Therefore, we do not find any evidence of tidal perturbation or disruption in this part of the SMC. 
    \item The arm-like and the arc-like features do not show differences in kinematics with respect to the surrounding regions. The arm-like and the arc-like structures are probably stellar over densities, which were formed at a similar period in time.
    \item In the SMC-Shell region, we do not detect any propagation in star formation. Most parts of the SMC-Shell experienced two episodes of star formation ranging between 40 to 400 Myr ago. The episode at $\sim$ 260 Myr is probably linked to the recent interaction between the MCs, whereas the youngest episode at $\sim$ 60 Myr could be due to the pericentric passage of the SMC around the MW. 
\end{enumerate}
\label{section:con}
\section*{Acknowledgements}
We want to thank the anonymous referee for the insightful review that helped in the improvement of the paper.
We thank Dr. Prasant Nayak and Abinaya O. Omkumar for their suggestions for this work. AS acknowledges support from SERB for the POWER fellowship. SS acknowledges support from the Science and Engineering Research Board of India through the Ramanujan Fellowship. This publication uses the data from the AstroSat mission of the ISRO, archived at the Indian Space Science Data Centre (ISSDC). The optical and PM data utilized in this work have been obtained from the European Space Agency (ESA) space mission Gaia (\href{https://www.cosmos.esa.int/gaia}{https://www.cosmos.esa.int/gaia}). We are grateful to the Gaia Data Processing and Analysis Consortium (DPAC, \href{https://www.cosmos.esa.int/web/gaia/dpac/consortium}{https://www.cosmos.esa.int/web/gaia/dpac/consortium}) for their ongoing efforts in processing Gaia data. The DPAC's work is made possible through funding provided by national institutions, with a special acknowledgment to the institutions participating in the Gaia MultiLateral Agreement (MLA). Through this research work, we have used the PYTHON packages, like NUMPY \citep{2020Natur.585..357H}, ASTROPY \citep{2013A&A...558A..33A}, MATPLOTLIB \citep{2007CSE.....9...90H} and SCIPY \citep{2020NatMe..17..261V}. 
 
\section*{Data Availability}
The data underlying this article are publicly available at \href{https://astrobrowse.issdc.gov.in/astro\_archive/archive/Home.jsp}{$https://astrobrowse.issdc.gov.in/astro\_archive/archive/Home.jsp$}



\begin{thebibliography}{}
\makeatletter
\relax
\def\mn@urlcharsother{\let\do\@makeother \do\$\do\&\do\#\do\^\do\_\do\%\do\~}
\def\mn@doi{\begingroup\mn@urlcharsother \@ifnextchar [ {\mn@doi@} {\mn@doi@[]}}
\def\mn@doi@[#1]#2{\def\@tempa{#1}\ifx\@tempa\@empty \href {http://dx.doi.org/#2} {doi:#2}\else \href {http://dx.doi.org/#2} {#1}\fi \endgroup}
\def\mn@eprint#1#2{\mn@eprint@#1:#2::\@nil}
\def\mn@eprint@arXiv#1{\href {http://arxiv.org/abs/#1} {{\tt arXiv:#1}}}
\def\mn@eprint@dblp#1{\href {http://dblp.uni-trier.de/rec/bibtex/#1.xml} {dblp:#1}}
\def\mn@eprint@#1:#2:#3:#4\@nil{\def\@tempa {#1}\def\@tempb {#2}\def\@tempc {#3}\ifx \@tempc \@empty \let \@tempc \@tempb \let \@tempb \@tempa \fi \ifx \@tempb \@empty \def\@tempb {arXiv}\fi \@ifundefined {mn@eprint@\@tempb}{\@tempb:\@tempc}{\expandafter \expandafter \csname mn@eprint@\@tempb\endcsname \expandafter{\@tempc}}}

\bibitem[\protect\citeauthoryear{{Albers}, {MacGillivray}, {Beard}  \& {Chromey}}{{Albers} et~al.}{1987}]{1987A&A...182L...8A}
{Albers} H.,  {MacGillivray} H.~T.,  {Beard} S.~M.,   {Chromey} F.~R.,  1987, \aap, \href {https://ui.adsabs.harvard.edu/abs/1987A&A...182L...8A} {182, L8}

\bibitem[\protect\citeauthoryear{{Astropy Collaboration} et~al.,}{{Astropy Collaboration} et~al.}{2013}]{2013A&A...558A..33A}
{Astropy Collaboration} et~al., 2013, \mn@doi [\aap] {10.1051/0004-6361/201322068}, \href {https://ui.adsabs.harvard.edu/abs/2013A&A...558A..33A} {558, A33}

\bibitem[\protect\citeauthoryear{Besla, Kallivayalil, Hernquist, Robertson, Cox, van~der Marel  \& Alcock}{Besla et~al.}{2007}]{besla2007magellanic}
Besla G.,  Kallivayalil N.,  Hernquist L.,  Robertson B.,  Cox T.,  van~der Marel R.~P.,   Alcock C.,  2007, The Astrophysical Journal, 668, 949

\bibitem[\protect\citeauthoryear{{Besla}, {Kallivayalil}, {Hernquist}, {van der Marel}, {Cox}  \& {Kere{\v{s}}}}{{Besla} et~al.}{2012}]{2012MNRAS.421.2109B}
{Besla} G.,  {Kallivayalil} N.,  {Hernquist} L.,  {van der Marel} R.~P.,  {Cox} T.~J.,   {Kere{\v{s}}} D.,  2012, \mn@doi [\mnras] {10.1111/j.1365-2966.2012.20466.x}, \href {https://ui.adsabs.harvard.edu/abs/2012MNRAS.421.2109B} {421, 2109}

\bibitem[\protect\citeauthoryear{{Bianchi}, {Shiao}  \& {Thilker}}{{Bianchi} et~al.}{2017}]{2017ApJS..230...24B}
{Bianchi} L.,  {Shiao} B.,   {Thilker} D.,  2017, \mn@doi [\apjs] {10.3847/1538-4365/aa7053}, \href {https://ui.adsabs.harvard.edu/abs/2017ApJS..230...24B} {230, 24}

\bibitem[\protect\citeauthoryear{{Bressan}, {Marigo}, {Girardi}, {Salasnich}, {Dal Cero}, {Rubele}  \& {Nanni}}{{Bressan} et~al.}{2012}]{2012MNRAS.427..127B}
{Bressan} A.,  {Marigo} P.,  {Girardi} L.,  {Salasnich} B.,  {Dal Cero} C.,  {Rubele} S.,   {Nanni} A.,  2012, \mn@doi [\mnras] {10.1111/j.1365-2966.2012.21948.x}, \href {https://ui.adsabs.harvard.edu/abs/2012MNRAS.427..127B} {427, 127}

\bibitem[\protect\citeauthoryear{{Brueck} \& {Marsoglu}}{{Brueck} \& {Marsoglu}}{1978}]{1978A&A....68..193B}
{Brueck} M.~T.,  {Marsoglu} A.,  1978, \aap, \href {https://ui.adsabs.harvard.edu/abs/1978A&A....68..193B} {68, 193}

\bibitem[\protect\citeauthoryear{{Chandra} et~al.,}{{Chandra} et~al.}{2023}]{2023arXiv230615719C}
{Chandra} V.,  et~al., 2023, \mn@doi [arXiv e-prints] {10.48550/arXiv.2306.15719}, \href {https://ui.adsabs.harvard.edu/abs/2023arXiv230615719C} {p. arXiv:2306.15719}

\bibitem[\protect\citeauthoryear{{Choi}, {Olsen}, {Besla}, {van der Marel}, {Zivick}, {Kallivayalil}  \& {Nidever}}{{Choi} et~al.}{2022}]{2022ApJ...927..153C}
{Choi} Y.,  {Olsen} K. A.~G.,  {Besla} G.,  {van der Marel} R.~P.,  {Zivick} P.,  {Kallivayalil} N.,   {Nidever} D.~L.,  2022, \mn@doi [\apj] {10.3847/1538-4357/ac4e90}, \href {https://ui.adsabs.harvard.edu/abs/2022ApJ...927..153C} {927, 153}

\bibitem[\protect\citeauthoryear{{D'Onghia} \& {Fox}}{{D'Onghia} \& {Fox}}{2016}]{2016ARA&A..54..363D}
{D'Onghia} E.,  {Fox} A.~J.,  2016, \mn@doi [\araa] {10.1146/annurev-astro-081915-023251}, \href {https://ui.adsabs.harvard.edu/abs/2016ARA&A..54..363D} {54, 363}

\bibitem[\protect\citeauthoryear{{De Leo}, {Carrera}, {No{\"e}l}, {Read}, {Erkal}  \& {Gallart}}{{De Leo} et~al.}{2020}]{2020MNRAS.495...98..De..Leo..tidal..radius}
{De Leo} M.,  {Carrera} R.,  {No{\"e}l} N. E.~D.,  {Read} J.~I.,  {Erkal} D.,   {Gallart} C.,  2020, \mn@doi [\mnras] {10.1093/mnras/staa1122}, \href {https://ui.adsabs.harvard.edu/abs/2020MNRAS.495...98D} {495, 98}

\bibitem[\protect\citeauthoryear{{Dias} et~al.,}{{Dias} et~al.}{2022}]{2022MNRAS.512.4334Das..D..tidal..radius}
{Dias} B.,  et~al., 2022, \mn@doi [\mnras] {10.1093/mnras/stac259}, \href {https://ui.adsabs.harvard.edu/abs/2022MNRAS.512.4334D} {512, 4334}

\bibitem[\protect\citeauthoryear{{Efremov}}{{Efremov}}{2002}]{2002ARep...46..791E}
{Efremov} Y.~N.,  2002, \mn@doi [Astronomy Reports] {10.1134/1.1515091}, \href {https://ui.adsabs.harvard.edu/abs/2002ARep...46..791E} {46, 791}

\bibitem[\protect\citeauthoryear{{El Youssoufi} et~al.,}{{El Youssoufi} et~al.}{2023}]{2023arXiv230414368E}
{El Youssoufi} D.,  et~al., 2023, \mn@doi [arXiv e-prints] {10.48550/arXiv.2304.14368}, \href {https://ui.adsabs.harvard.edu/abs/2023arXiv230414368E} {p. arXiv:2304.14368}

\bibitem[\protect\citeauthoryear{{Gaia Collaboration} et~al.,}{{Gaia Collaboration} et~al.}{2018a}]{2018A&A...616A...1G}
{Gaia Collaboration} et~al., 2018a, \mn@doi [\aap] {10.1051/0004-6361/201833051}, \href {https://ui.adsabs.harvard.edu/abs/2018A&A...616A...1G} {616, A1}

\bibitem[\protect\citeauthoryear{{Gaia Collaboration} et~al.,}{{Gaia Collaboration} et~al.}{2018b}]{2018A&A...616A..12G}
{Gaia Collaboration} et~al., 2018b, \mn@doi [\aap] {10.1051/0004-6361/201832698}, \href {https://ui.adsabs.harvard.edu/abs/2018A&A...616A..12G} {616, A12}

\bibitem[\protect\citeauthoryear{{Gaia Collaboration} et~al.,}{{Gaia Collaboration} et~al.}{2021}]{2021A&A...649A...1G}
{Gaia Collaboration} et~al., 2021, \mn@doi [\aap] {10.1051/0004-6361/202039657}, \href {https://ui.adsabs.harvard.edu/abs/2021A&A...649A...1G} {649, A1}

\bibitem[\protect\citeauthoryear{{Gaia Collaboration} et~al.,}{{Gaia Collaboration} et~al.}{2022}]{2022arXiv220800211G}
{Gaia Collaboration} et~al., 2022, \mn@doi [arXiv e-prints] {10.48550/arXiv.2208.00211}, \href {https://ui.adsabs.harvard.edu/abs/2022arXiv220800211G} {p. arXiv:2208.00211}

\bibitem[\protect\citeauthoryear{{Gaia Collaboration} et~al.,}{{Gaia Collaboration} et~al.}{2023}]{2023A&A...674A...1G}
{Gaia Collaboration} et~al., 2023, \mn@doi [\aap] {10.1051/0004-6361/202243940}, \href {https://ui.adsabs.harvard.edu/abs/2023A&A...674A...1G} {674, A1}

\bibitem[\protect\citeauthoryear{{Harris} \& {Zaritsky}}{{Harris} \& {Zaritsky}}{2004}]{2004AJ....127.1531H}
{Harris} J.,  {Zaritsky} D.,  2004, \mn@doi [\aj] {10.1086/381953}, \href {https://ui.adsabs.harvard.edu/abs/2004AJ....127.1531H} {127, 1531}

\bibitem[\protect\citeauthoryear{{Harris} et~al.,}{{Harris} et~al.}{2020}]{2020Natur.585..357H}
{Harris} C.~R.,  et~al., 2020, \mn@doi [\nat] {10.1038/s41586-020-2649-2}, \href {https://ui.adsabs.harvard.edu/abs/2020Natur.585..357H} {585, 357}

\bibitem[\protect\citeauthoryear{{Haschke}, {Grebel}  \& {Duffau}}{{Haschke} et~al.}{2011}]{2011AJ....141..158H}
{Haschke} R.,  {Grebel} E.~K.,   {Duffau} S.,  2011, \mn@doi [\aj] {10.1088/0004-6256/141/5/158}, \href {https://ui.adsabs.harvard.edu/abs/2011AJ....141..158H} {141, 158}

\bibitem[\protect\citeauthoryear{Hindman, Kerr  \& McGee}{Hindman et~al.}{1963}]{Hindman1963ALR}
Hindman J.~V.,  Kerr F.~J.,   McGee R.~X.,  1963, Australian Journal of Physics, 16, 570

\bibitem[\protect\citeauthoryear{{Hunter}}{{Hunter}}{2007}]{2007CSE.....9...90H}
{Hunter} J.~D.,  2007, \mn@doi [Computing in Science and Engineering] {10.1109/MCSE.2007.55}, \href {https://ui.adsabs.harvard.edu/abs/2007CSE.....9...90H} {9, 90}

\bibitem[\protect\citeauthoryear{{Indu} \& {Subramaniam}}{{Indu} \& {Subramaniam}}{2011}]{2011A&A...535A.115I}
{Indu} G.,  {Subramaniam} A.,  2011, \mn@doi [\aap] {10.1051/0004-6361/201117298}, \href {https://ui.adsabs.harvard.edu/abs/2011A&A...535A.115I} {535, A115}

\bibitem[\protect\citeauthoryear{{Irwin}, {Kunkel}  \& {Demers}}{{Irwin} et~al.}{1985}]{1985Natur.318..160I}
{Irwin} M.~J.,  {Kunkel} W.~E.,   {Demers} S.,  1985, \mn@doi [\nat] {10.1038/318160a0}, \href {https://ui.adsabs.harvard.edu/abs/1985Natur.318..160I} {318, 160}

\bibitem[\protect\citeauthoryear{{Joshi}, {Prasad Mohanty}  \& {Joshi}}{{Joshi} et~al.}{2016}]{2016RAA....16...61J}
{Joshi} Y.~C.,  {Prasad Mohanty} A.,   {Joshi} S.,  2016, \mn@doi [Research in Astronomy and Astrophysics] {10.1088/1674-4527/16/4/061}, \href {https://ui.adsabs.harvard.edu/abs/2016RAA....16...61J} {16, 61}

\bibitem[\protect\citeauthoryear{{Kallivayalil}, {van der Marel}  \& {Alcock}}{{Kallivayalil} et~al.}{2006}]{2006ApJ...652.1213K}
{Kallivayalil} N.,  {van der Marel} R.~P.,   {Alcock} C.,  2006, \mn@doi [\apj] {10.1086/508014}, \href {https://ui.adsabs.harvard.edu/abs/2006ApJ...652.1213K} {652, 1213}

\bibitem[\protect\citeauthoryear{{Kumar} et~al.,}{{Kumar} et~al.}{2012}]{2012SPIE.8443E..1NK}
{Kumar} A.,  et~al., 2012, in {Takahashi} T.,  {Murray} S.~S.,   {den Herder} J.-W.~A.,  eds,  Society of Photo-Optical Instrumentation Engineers (SPIE) Conference Series Vol. 8443, Space Telescopes and Instrumentation 2012: Ultraviolet to Gamma Ray. p. 84431N (\mn@eprint {arXiv} {1208.4670}), \mn@doi{10.1117/12.924507}

\bibitem[\protect\citeauthoryear{{Lemasle} et~al.,}{{Lemasle} et~al.}{2017}]{2017A&A...608A..85L}
{Lemasle} B.,  et~al., 2017, \mn@doi [\aap] {10.1051/0004-6361/201731370}, \href {https://ui.adsabs.harvard.edu/abs/2017A&A...608A..85L} {608, A85}

\bibitem[\protect\citeauthoryear{{Lin}, {Jones}  \& {Klemola}}{{Lin} et~al.}{1995}]{1995ApJ...439..652L}
{Lin} D.~N.~C.,  {Jones} B.~F.,   {Klemola} A.~R.,  1995, \mn@doi [\apj] {10.1086/175205}, \href {https://ui.adsabs.harvard.edu/abs/1995ApJ...439..652L} {439, 652}

\bibitem[\protect\citeauthoryear{Lindegren}{Lindegren}{2018}]{LL:LL-124}
Lindegren L.,  2018, {R}e-normalising the astrometric chi-square in {G}aia {D}{R}2, GAIA-C3-TN-LU-LL-124, \url {http://www.rssd.esa.int/doc_fetch.php?id=3757412}

\bibitem[\protect\citeauthoryear{{Lucchini}, {D'Onghia}, {Fox}, {Bustard}, {Bland-Hawthorn}  \& {Zweibel}}{{Lucchini} et~al.}{2020}]{2020Natur.585..203L}
{Lucchini} S.,  {D'Onghia} E.,  {Fox} A.~J.,  {Bustard} C.,  {Bland-Hawthorn} J.,   {Zweibel} E.,  2020, \mn@doi [\nat] {10.1038/s41586-020-2663-4}, \href {https://ui.adsabs.harvard.edu/abs/2020Natur.585..203L} {585, 203}

\bibitem[\protect\citeauthoryear{{Martin} et~al.,}{{Martin} et~al.}{2005}]{2005ApJ...619L...1M}
{Martin} D.~C.,  et~al., 2005, \mn@doi [\apjl] {10.1086/426387}, \href {https://ui.adsabs.harvard.edu/abs/2005ApJ...619L...1M} {619, L1}

\bibitem[\protect\citeauthoryear{{Mart{\'\i}nez-Delgado} et~al.,}{{Mart{\'\i}nez-Delgado} et~al.}{2019}]{2019A&A...631A..98M}
{Mart{\'\i}nez-Delgado} D.,  et~al., 2019, \mn@doi [\aap] {10.1051/0004-6361/201936021}, \href {https://ui.adsabs.harvard.edu/abs/2019A&A...631A..98M} {631, A98}

\bibitem[\protect\citeauthoryear{{Massana} et~al.,}{{Massana} et~al.}{2020}]{2020MNRAS.498.1034M}
{Massana} P.,  et~al., 2020, \mn@doi [\mnras] {10.1093/mnras/staa2451}, \href {https://ui.adsabs.harvard.edu/abs/2020MNRAS.498.1034M} {498, 1034}

\bibitem[\protect\citeauthoryear{{Mathewson}}{{Mathewson}}{1985}]{1985PASA....6..104M}
{Mathewson} D.~S.,  1985, \mn@doi [\pasa] {10.1017/S1323358000026771}, \href {https://ui.adsabs.harvard.edu/abs/1985PASA....6..104M} {6, 104}

\bibitem[\protect\citeauthoryear{{Miller} et~al.,}{{Miller} et~al.}{2022}]{2022MNRAS.512.1196M}
{Miller} A.~E.,  et~al., 2022, \mn@doi [\mnras] {10.1093/mnras/stac508}, \href {https://ui.adsabs.harvard.edu/abs/2022MNRAS.512.1196M} {512, 1196}

\bibitem[\protect\citeauthoryear{{Nidever}, {Majewski}  \& {Butler Burton}}{{Nidever} et~al.}{2008}]{2008ApJ...679..432N}
{Nidever} D.~L.,  {Majewski} S.~R.,   {Butler Burton} W.,  2008, \mn@doi [\apj] {10.1086/587042}, \href {https://ui.adsabs.harvard.edu/abs/2008ApJ...679..432N} {679, 432}

\bibitem[\protect\citeauthoryear{{Nidever} et~al.,}{{Nidever} et~al.}{2017}]{2017AJ....154..199N}
{Nidever} D.~L.,  et~al., 2017, \mn@doi [\aj] {10.3847/1538-3881/aa8d1c}, \href {https://ui.adsabs.harvard.edu/abs/2017AJ....154..199N} {154, 199}

\bibitem[\protect\citeauthoryear{{Niederhofer} et~al.,}{{Niederhofer} et~al.}{2021}]{2021MNRAS.502.2859N}
{Niederhofer} F.,  et~al., 2021, \mn@doi [\mnras] {10.1093/mnras/stab206}, \href {https://ui.adsabs.harvard.edu/abs/2021MNRAS.502.2859N} {502, 2859}

\bibitem[\protect\citeauthoryear{{No{\"e}l}, {Gallart}, {Costa}  \& {M{\'e}ndez}}{{No{\"e}l} et~al.}{2007}]{2007AJ....133.2037N}
{No{\"e}l} N. E.~D.,  {Gallart} C.,  {Costa} E.,   {M{\'e}ndez} R.~A.,  2007, \mn@doi [\aj] {10.1086/512668}, \href {https://ui.adsabs.harvard.edu/abs/2007AJ....133.2037N} {133, 2037}

\bibitem[\protect\citeauthoryear{{No{\"e}l}, {Aparicio}, {Gallart}, {Hidalgo}, {Costa}  \& {M{\'e}ndez}}{{No{\"e}l} et~al.}{2009}]{2009ApJ...705.1260N}
{No{\"e}l} N. E.~D.,  {Aparicio} A.,  {Gallart} C.,  {Hidalgo} S.~L.,  {Costa} E.,   {M{\'e}ndez} R.~A.,  2009, \mn@doi [\apj] {10.1088/0004-637X/705/2/1260}, \href {https://ui.adsabs.harvard.edu/abs/2009ApJ...705.1260N} {705, 1260}

\bibitem[\protect\citeauthoryear{{Piatti}}{{Piatti}}{2014}]{2014..Piatti..Age}
{Piatti} A.~E.,  2014, \mn@doi [\mnras] {10.1093/mnras/stu534}, \href {https://ui.adsabs.harvard.edu/abs/2014MNRAS.440.3091P} {440, 3091}

\bibitem[\protect\citeauthoryear{{Piatti}}{{Piatti}}{2022}]{2022MNRAS.509.3462P}
{Piatti} A.~E.,  2022, \mn@doi [\mnras] {10.1093/mnras/stab3190}, \href {https://ui.adsabs.harvard.edu/abs/2022MNRAS.509.3462P} {509, 3462}

\bibitem[\protect\citeauthoryear{{Postma} \& {Leahy}}{{Postma} \& {Leahy}}{2017}]{2017PASP..129k5002P}
{Postma} J.~E.,  {Leahy} D.,  2017, \mn@doi [\pasp] {10.1088/1538-3873/aa8800}, \href {https://ui.adsabs.harvard.edu/abs/2017PASP..129k5002P} {129, 115002}

\bibitem[\protect\citeauthoryear{{Postma} \& {Leahy}}{{Postma} \& {Leahy}}{2021}]{2021JApA...42...30P}
{Postma} J.~E.,  {Leahy} D.,  2021, \mn@doi [Journal of Astrophysics and Astronomy] {10.1007/s12036-020-09689-w}, \href {https://ui.adsabs.harvard.edu/abs/2021JApA...42...30P} {42, 30}

\bibitem[\protect\citeauthoryear{{Romaniello} et~al.,}{{Romaniello} et~al.}{2008}]{2008A&A...488..731R}
{Romaniello} M.,  et~al., 2008, \mn@doi [\aap] {10.1051/0004-6361:20065661}, \href {https://ui.adsabs.harvard.edu/abs/2008A&A...488..731R} {488, 731}

\bibitem[\protect\citeauthoryear{{Russell} \& {Dopita}}{{Russell} \& {Dopita}}{1992}]{1992ApJ...384..508R}
{Russell} S.~C.,  {Dopita} M.~A.,  1992, \mn@doi [\apj] {10.1086/170893}, \href {https://ui.adsabs.harvard.edu/abs/1992ApJ...384..508R} {384, 508}

\bibitem[\protect\citeauthoryear{{Sahu} et~al.,}{{Sahu} et~al.}{2022}]{2022MNRAS.514.1122S}
{Sahu} S.,  et~al., 2022, \mn@doi [\mnras] {10.1093/mnras/stac1209}, \href {https://ui.adsabs.harvard.edu/abs/2022MNRAS.514.1122S} {514, 1122}

\bibitem[\protect\citeauthoryear{{Sakowska}, {No{\"e}l}, {Ruiz-Lara}  \& {Gallart}}{{Sakowska} et~al.}{2023}]{2023sakowska}
{Sakowska} J.~D.,  {No{\"e}l} N. E.~D.,  {Ruiz-Lara} T.,   {Gallart} C.,  2023, \mn@doi [arXiv e-prints] {10.48550/arXiv.2305.02755}, \href {https://ui.adsabs.harvard.edu/abs/2023arXiv230502755S} {p. arXiv:2305.02755}

\bibitem[\protect\citeauthoryear{{Salem}, {Besla}, {Bryan}, {Putman}, {van der Marel}  \& {Tonnesen}}{{Salem} et~al.}{2015}]{2015ApJ...815...77S}
{Salem} M.,  {Besla} G.,  {Bryan} G.,  {Putman} M.,  {van der Marel} R.~P.,   {Tonnesen} S.,  2015, \mn@doi [\apj] {10.1088/0004-637X/815/1/77}, \href {https://ui.adsabs.harvard.edu/abs/2015ApJ...815...77S} {815, 77}

\bibitem[\protect\citeauthoryear{{Stetson}}{{Stetson}}{1987}]{1987PASP...99..191S}
{Stetson} P.~B.,  1987, \mn@doi [\pasp] {10.1086/131977}, \href {https://ui.adsabs.harvard.edu/abs/1987PASP...99..191S} {99, 191}

\bibitem[\protect\citeauthoryear{Subramaniam et~al.,}{Subramaniam et~al.}{2016}]{subramaniam2016orbit}
Subramaniam A.,  et~al., 2016, in Space Telescopes and Instrumentation 2016: Ultraviolet to Gamma Ray. pp 399--408

\bibitem[\protect\citeauthoryear{{Sun} et~al.,}{{Sun} et~al.}{2017}]{2017ApJ...849..149S}
{Sun} N.-C.,  et~al., 2017, \mn@doi [\apj] {10.3847/1538-4357/aa911e}, \href {https://ui.adsabs.harvard.edu/abs/2017ApJ...849..149S} {849, 149}

\bibitem[\protect\citeauthoryear{{Sun} et~al.,}{{Sun} et~al.}{2018}]{2018ApJ...858...31S}
{Sun} N.-C.,  et~al., 2018, \mn@doi [\apj] {10.3847/1538-4357/aabc50}, \href {https://ui.adsabs.harvard.edu/abs/2018ApJ...858...31S} {858, 31}

\bibitem[\protect\citeauthoryear{{Tandon} et~al.,}{{Tandon} et~al.}{2017a}]{2017JApA...38...28T}
{Tandon} S.~N.,  et~al., 2017a, \mn@doi [Journal of Astrophysics and Astronomy] {10.1007/s12036-017-9445-x}, \href {https://ui.adsabs.harvard.edu/abs/2017JApA...38...28T} {38, 28}

\bibitem[\protect\citeauthoryear{{Tandon} et~al.,}{{Tandon} et~al.}{2017b}]{2017AJ....154..128T}
{Tandon} S.~N.,  et~al., 2017b, \mn@doi [\aj] {10.3847/1538-3881/aa8451}, \href {https://ui.adsabs.harvard.edu/abs/2017AJ....154..128T} {154, 128}

\bibitem[\protect\citeauthoryear{{Tandon} et~al.,}{{Tandon} et~al.}{2020}]{2020AJ....159..158T}
{Tandon} S.~N.,  et~al., 2020, \mn@doi [\aj] {10.3847/1538-3881/ab72a3}, \href {https://ui.adsabs.harvard.edu/abs/2020AJ....159..158T} {159, 158}

\bibitem[\protect\citeauthoryear{Tatton et~al.,}{Tatton et~al.}{2021}]{10.1093/mnras/staa3857}
Tatton B.~L.,  et~al., 2021, \mn@doi [Monthly Notices of the Royal Astronomical Society] {10.1093/mnras/staa3857}, 504, 2983

\bibitem[\protect\citeauthoryear{{Thilker} et~al.,}{{Thilker} et~al.}{2007}]{2007ApJS..173..538T}
{Thilker} D.~A.,  et~al., 2007, \mn@doi [\apjs] {10.1086/523853}, \href {https://ui.adsabs.harvard.edu/abs/2007ApJS..173..538T} {173, 538}

\bibitem[\protect\citeauthoryear{{Vasiliev}}{{Vasiliev}}{2023}]{2023arXiv230604837V}
{Vasiliev} E.,  2023, \mn@doi [arXiv e-prints] {10.48550/arXiv.2306.04837}, \href {https://ui.adsabs.harvard.edu/abs/2023arXiv230604837V} {p. arXiv:2306.04837}

\bibitem[\protect\citeauthoryear{{Virtanen} et~al.,}{{Virtanen} et~al.}{2020}]{2020NatMe..17..261V}
{Virtanen} P.,  et~al., 2020, \mn@doi [Nature Methods] {10.1038/s41592-019-0686-2}, \href {https://ui.adsabs.harvard.edu/abs/2020NatMe..17..261V} {17, 261}

\bibitem[\protect\citeauthoryear{{Zaritsky}, {Harris}, {Grebel}  \& {Thompson}}{{Zaritsky} et~al.}{2000}]{2000ApJ...534L..53Z}
{Zaritsky} D.,  {Harris} J.,  {Grebel} E.~K.,   {Thompson} I.~B.,  2000, \mn@doi [\apjl] {10.1086/312649}, \href {https://ui.adsabs.harvard.edu/abs/2000ApJ...534L..53Z} {534, L53}

\bibitem[\protect\citeauthoryear{{Zivick} et~al.,}{{Zivick} et~al.}{2018}]{2018ApJ...864...55Z}
{Zivick} P.,  et~al., 2018, \mn@doi [\apj] {10.3847/1538-4357/aad4b0}, \href {https://ui.adsabs.harvard.edu/abs/2018ApJ...864...55Z} {864, 55}

\bibitem[\protect\citeauthoryear{{de Grijs} \& {Bono}}{{de Grijs} \& {Bono}}{2015}]{2015AJ....149..179D}
{de Grijs} R.,  {Bono} G.,  2015, \mn@doi [\aj] {10.1088/0004-6256/149/6/179}, \href {https://ui.adsabs.harvard.edu/abs/2015AJ....149..179D} {149, 179}

\bibitem[\protect\citeauthoryear{{de Grijs}, {Wicker}  \& {Bono}}{{de Grijs} et~al.}{2014}]{2014AJ....147..122D}
{de Grijs} R.,  {Wicker} J.~E.,   {Bono} G.,  2014, \mn@doi [\aj] {10.1088/0004-6256/147/5/122}, \href {https://ui.adsabs.harvard.edu/abs/2014AJ....147..122D} {147, 122}

\bibitem[\protect\citeauthoryear{{de Vaucouleurs} \& {Freeman}}{{de Vaucouleurs} \& {Freeman}}{1972}]{1972VA.....14..163D}
{de Vaucouleurs} G.,  {Freeman} K.~C.,  1972, \mn@doi [Vistas in Astronomy] {10.1016/0083-6656(72)90026-8}, \href {https://ui.adsabs.harvard.edu/abs/1972VA.....14..163D} {14, 163}

\bibitem[\protect\citeauthoryear{{van der Marel} \& {Cioni}}{{van der Marel} \& {Cioni}}{2001}]{2001AJ....122.1807V}
{van der Marel} R.~P.,  {Cioni} M.-R.~L.,  2001, \mn@doi [\aj] {10.1086/323099}, \href {https://ui.adsabs.harvard.edu/abs/2001AJ....122.1807V} {122, 1807}

\bibitem[\protect\citeauthoryear{van~der Marel, Kallivayalil  \& Besla}{van~der Marel et~al.}{2008}]{van_der_Marel_2008}
van~der Marel R.~P.,  Kallivayalil N.,   Besla G.,  2008, \mn@doi [Proceedings of the International Astronomical Union] {10.1017/s1743921308028299}, 4, 81

\makeatother
\end{thebibliography}

\appendix

\section{List of Caveats}
\label{append: appendix}
\begin{itemize}
    \item In this study, the region around the Galactic globular cluster NGC 362 is excluded as this region is contaminated with the Milky Way stars. We also point out that the UVIT has not observed the SMC-Shell at larger radii, which limits our analysis of the NE extent of the inner SMC (the kinematic edge and photometric edge).

    \item The SMC-Shell region shows continuous star formation, and to estimate the age range of star formation, we have visually overlaid isochrones on the FUV--optical CMDs. Even though the quoted age values may be subjective due to the visual overlay of isochrones, one can still notice deviation if the shift in the age of the isochrone is beyond 10\% error of the estimated age (shown in Fig.~\ref{fig:arm_appendix}).
    \begin{figure}  
          \centering
          \includegraphics[width=\columnwidth,trim={0 0 0 0},clip]{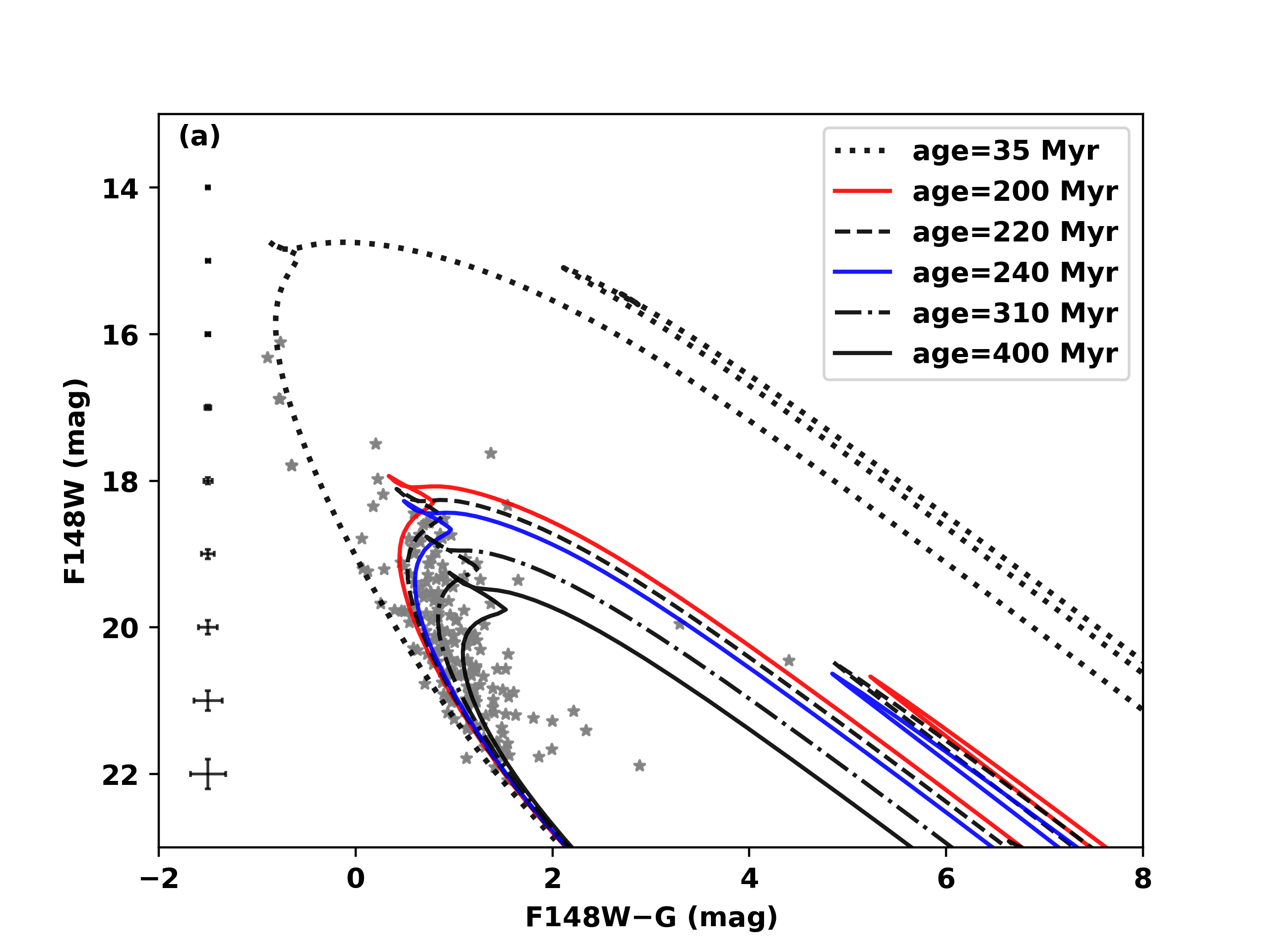}
          \\
           \includegraphics[width=\columnwidth,trim={0 0 0 1.0cm},clip]{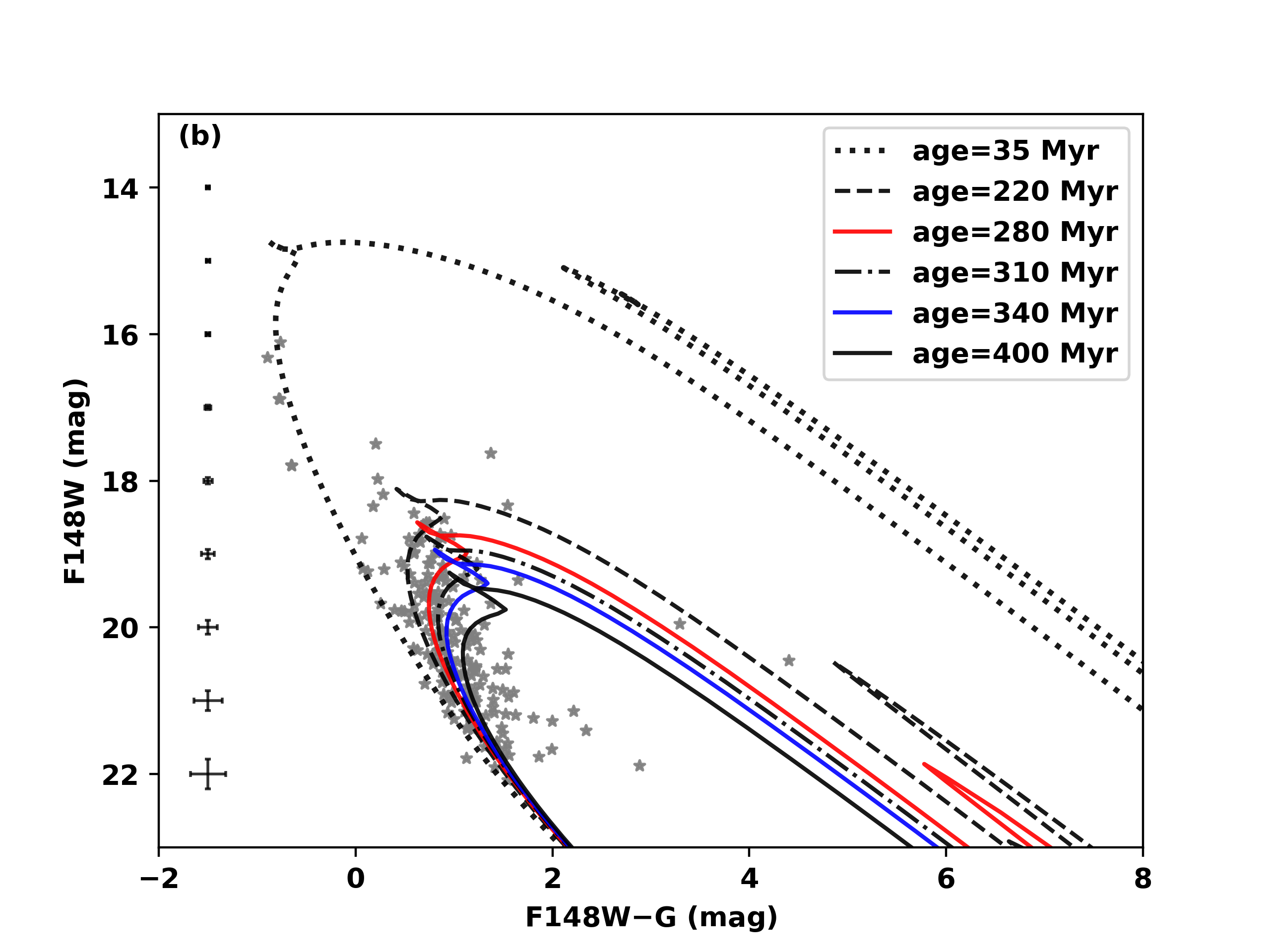}
        
          \caption{(a) and (b) show the FUV--optical CMDs of the arm-like structure (Fig.~\ref{fig:parm}b) displaying $\sim$ 10\% error range for ages 220 Myr and 310 Myr, respectively.} \label{fig:arm_appendix}
    \end{figure}

    \item We limit our upper age estimation at 400 Myr as stars older than that have an FUV magnitude fainter than 20 mag, which is associated with the larger photometric error (Fig.\ref{fig:Mag_Mage} and \ref{fig:cmd}a).
\end{itemize}

\label{lastpage}
\end{document}